\documentclass[twoside,twocolumn,9pt]{article}
\usepackage{extsizes}
\usepackage[super,sort&compress,comma]{natbib} 
\usepackage[version=3]{mhchem}
\usepackage[left=1.5cm, right=1.5cm, top=1.785cm, bottom=2.0cm]{geometry}
\usepackage{balance}
\usepackage{widetext}
\usepackage{times,mathptmx}
\usepackage{sectsty}
\usepackage{graphicx} 
\usepackage{lastpage}
\usepackage[format=plain,justification=raggedright,singlelinecheck=false,font={stretch=1.125,small,sf},labelfont=bf,labelsep=space]{caption}
\usepackage{float}
\usepackage{fancyhdr}
\usepackage{fnpos}
\usepackage[english]{babel}
\usepackage{array}
\usepackage{droidsans}
\usepackage{charter}
\usepackage[T1]{fontenc}
\usepackage[usenames,dvipsnames]{xcolor}
\usepackage{setspace}
\usepackage[compact]{titlesec}

\usepackage{epstopdf}

\definecolor{cream}{RGB}{222,217,201}

\begin{document}

\pagestyle{fancy}
\thispagestyle{plain}
\fancypagestyle{plain}{

\renewcommand{\headrulewidth}{0pt}
}

\makeFNbottom
\makeatletter
\renewcommand\LARGE{\@setfontsize\LARGE{15pt}{17}}
\renewcommand\Large{\@setfontsize\Large{12pt}{14}}
\renewcommand\large{\@setfontsize\large{10pt}{12}}
\renewcommand\footnotesize{\@setfontsize\footnotesize{7pt}{10}}
\makeatother

\renewcommand{\thefootnote}{\fnsymbol{footnote}}
\renewcommand\footnoterule{\vspace*{1pt}%
\color{cream}\hrule width 3.5in height 0.4pt \color{black}\vspace*{5pt}} 
\setcounter{secnumdepth}{5}

\makeatletter 
\renewcommand\@biblabel[1]{#1}            
\renewcommand\@makefntext[1]%
{\noindent\makebox[0pt][r]{\@thefnmark\,}#1}
\makeatother 
\renewcommand{\figurename}{\small{Fig.}~}
\sectionfont{\sffamily\Large}
\subsectionfont{\normalsize}
\subsubsectionfont{\bf}
\setstretch{1.125} 
\setlength{\skip\footins}{0.8cm}
\setlength{\footnotesep}{0.25cm}
\setlength{\jot}{10pt}
\titlespacing*{\section}{0pt}{4pt}{4pt}
\titlespacing*{\subsection}{0pt}{15pt}{1pt}

\fancyfoot{}
\fancyfoot[LO,RE]{\vspace{-7.1pt}}
\fancyfoot[CO]{\vspace{-7.1pt}\hspace{13.2cm}}
\fancyfoot[CE]{\vspace{-7.2pt}\hspace{-14.2cm}}
\fancyfoot[RO]{\footnotesize{\sffamily{1--\pageref{LastPage} ~\textbar  \hspace{2pt}\thepage}}}
\fancyfoot[LE]{\footnotesize{\sffamily{\thepage~\textbar\hspace{3.45cm} 1--\pageref{LastPage}}}}
\fancyhead{}
\renewcommand{\headrulewidth}{0pt} 
\renewcommand{\footrulewidth}{0pt}
\setlength{\arrayrulewidth}{1pt}
\setlength{\columnsep}{6.5mm}
\setlength\bibsep{1pt}

\makeatletter 
\newlength{\figrulesep} 
\setlength{\figrulesep}{0.5\textfloatsep} 

\newcommand{\topfigrule}{\vspace*{-1pt}%
\noindent{\color{cream}\rule[-\figrulesep]{\columnwidth}{1.5pt}} }

\newcommand{\botfigrule}{\vspace*{-2pt}%
\noindent{\color{cream}\rule[\figrulesep]{\columnwidth}{1.5pt}} }

\newcommand{\dblfigrule}{\vspace*{-1pt}%
\noindent{\color{cream}\rule[-\figrulesep]{\textwidth}{1.5pt}} }

\makeatother

\twocolumn[
  \begin{@twocolumnfalse}
\vspace{3cm}
\sffamily
\LARGE{\textbf{Machine learning enables long time scale molecular photodynamics simulations}} \\
\vspace{0.3cm} \\

  \noindent\large{Julia Westermayr,\textit{$^{a}$} Michael Gastegger,\textit{$^{b}$} Maximilian F. S. J. Menger,\textit{$^{a,c}$} Sebastian Mai,\textit{$^{a}$} 
  Leticia Gonz\'{a}lez\textit{$^{a}$} and Philipp Marquetand$^{\ast}$\textit{$^{a}$}} \\

 \noindent\normalsize{Photo-induced processes are fundamental in nature but accurate simulations are seriously limited by the cost of the underlying quantum chemical calculations, hampering their application for long time scales. Here we introduce a method based on machine learning to overcome this bottleneck and enable accurate photodynamics on nanosecond time scales, which are otherwise out of reach with contemporary approaches. Instead of expensive quantum chemistry during molecular dynamics simulations, we use deep neural networks to learn the relationship between a molecular geometry and its high-dimensional electronic properties. As an example, the time evolution of the methylenimmonium cation for one nanosecond is used to demonstrate that machine learning algorithms can outperform standard excited-state molecular dynamics approaches in their computational efficiency while delivering the same accuracy.} \\

 \end{@twocolumnfalse} \vspace{0.6cm}

  ]

\renewcommand*\rmdefault{bch}\normalfont\upshape
\rmfamily
\section*{}
\vspace{-1cm}


\footnotetext{\textit{$^{a}$~Institute of Theoretical Chemistry, Department of Chemistry, University of Vienna, 1090 Vienna, Austria. }}
\footnotetext{\textit{$^{b}$~Machine Learning Group, Technical University of Berlin, 10587 Berlin, Germany. }}
\footnotetext{\textit{$^{c}$~Dipartimento di Chimica e Chimica Industriale, University of Pisa, Via G. Moruzzi 13, 56124 Pisa, Italy. }}
\footnotetext{\textit{$^{\ast}$~ Corresponding author. E-mail: philipp.marquetand@univie.ac.at }}
\footnotetext{Electronic Supplementary Information (ESI) available at the end of the document}


\section{Introduction}

Machine learning (ML) is revolutionizing the most diverse domains, like image recognition,\cite{Goodfellow2016} playing board games,\cite{Silver2016N} or society integration of refugees.\cite{Bansak2018S} Also in chemistry, an increasing range of applications is being tackled with ML, for example, the design and discovery of new molecules and materials.\cite{Sanchez-Lengeling2018S,Butler2018N,Goldsmith2018AJ} In the present study, we show how ML enables efficient photodynamics simulations. 
Photodynamics is the study of photo-induced processes that occur after a molecule is exposed to light. Photosynthesis or DNA photodamage leading to skin cancer are only two examples of phenomena that involve molecules interacting with light.\cite{Cerullo2002S,Schultz2004S,Schreier2007S,Rauer2016JACS,Romero2017N} The simulation of such processes has been key to learn structure-dynamics-function relationships that can be used to guide the design of photonic materials, such as photosensitive drugs,\cite{Ahmad2016IJP} photocatalysts\cite{Sanchez-Lengeling2018S} and photovoltaics.\cite{Mathew2014NC,Bartok2017SA}

\begin{figure}[h]
\centering
  \includegraphics[height=8.5cm]{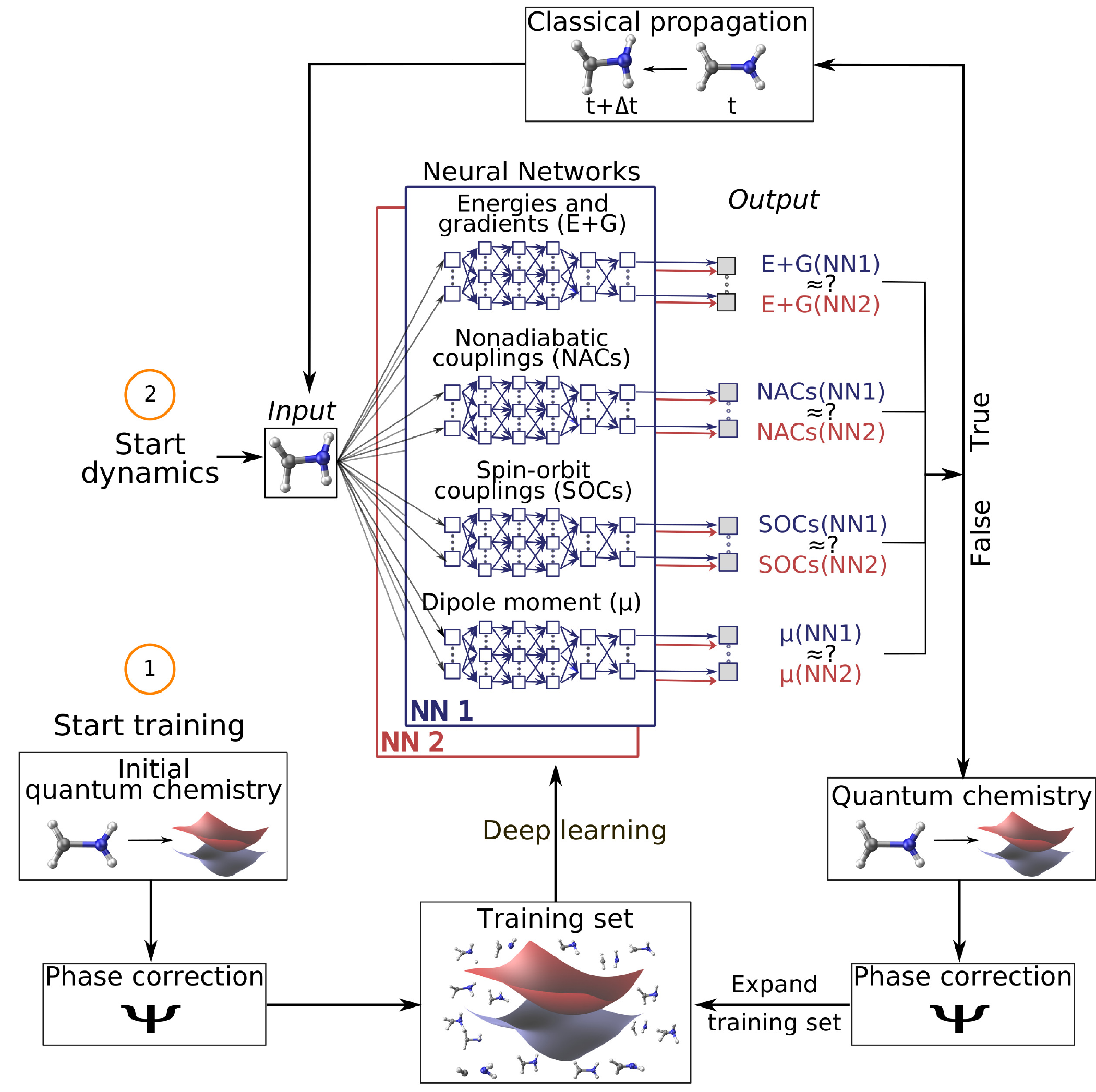}
  \caption{Schematic workflow of surface hopping molecular dynamics with deep NNs:  The scheme starts from a set of initial quantum chemical calculations, which are pre-processed by a phase-correction algorithm and constitute an initial training set. Using this set, two deep NNs (NN1 and NN2) are trained and replace the quantum chemical calculations of energies (E) and gradients (G), nonadiabatic couplings (NACs), spin-orbit couplings (SOCs) and dipole moments ($\mu$). The dynamics calculation starts with an input geometry, for which the two NNs provide all electronic quantities. If the outcome of both NNs is sufficiently similar, the configurational space around this input geometry is adequately represented by the training set and the electronic quantities are used for a propagation time step. If not, the nuclear configuration is recomputed with quantum chemistry, phase corrected and included in the training set -- a process referred to as adaptive sampling.  The NNs are then re-trained and a new dynamics cycle is started.}
  \label{fig:1}
\end{figure}

Computer simulations of photodynamics typically rely on molecular dynamics simulations of coupled nuclei and electrons. These simulations require the computation of high-dimensional potential energy surfaces (PESs), i.e., the electronic energy levels of the molecule for all possible molecular configurations, using quantum chemistry. The calculation of these PESs is usually the most expensive part of the dynamics simulations\cite{Mai2018WCMS} and therefore, different approximations are necessary and ubiquitous. For the electronic ground state, the time-consuming quantum chemical calculations are often replaced with force fields\cite{Chmiela2018NC} but no force fields are available to describe electronically excited states. Another drawback of most conventional force fields is their inability to describe the breaking and formation of chemical bonds. Recently, increasing effort has been devoted to ML potentials,\cite{Rupp2015IJQC,Lilienfeld2018ACIE} where an accurate representation of the ground state PES including bond breaking and formation is promised.\cite{Bartok2010PRL,Li2015PRL,Rupp2015JPCL,Behler2016JCP,Gastegger2017CS,Botu2017JPCC,Smith2017CS,Behler2017ACIE,Zong2018npjCM,Bartok2018PRX,Xia2018NC,Chmiela2018NC,Chan2019JPCC,Christensen2019JCP} Similarly, modified Shepard interpolation is used to construct PESs in low-dimensional systems and adapt them in out-of-confidence region.\cite{Netzloff2006JCP,Bettens1999JCP} However, the problem of obtaining accurate full dimensional PESs for excited states in order to simulate long time photodynamics has not been solved yet. A few studies focused on the prediction of excited state dynamics as well as on excited-state properties such as spectral densities with ML.\cite{Behler2008PRB,Carbogno2010PRB,Haese2016CS,Liu2017SR,Hu2018JPCL,Dral2018JPCL,Chen2018JPCL,William2018JCP,Xie2018JCP,Guan2019PCCP} 
The breakdown of the Born-Oppenheimer approximation, leading to critical regions in the coupled excited state PESs\cite{Domcke2004} pose yet another obstacle to quantum chemistry (QC) and consequently also ML.\cite{Hu2018JPCL,Dral2018JPCL,Chen2018JPCL} Among those critical regions are conical intersections (or state crossings), where two PESs get into close proximity. The underlying elements that become important in such areas are nonadiabatic couplings (or spin-orbit couplings). They induce non-radiative transitions between two electronic states of the same (or different) spin-multiplicities involving ultrafast rearrangements of both nuclei and electrons. These challenges led to the need of intermittent quantum chemistry calculations\cite{Hu2018JPCL,Dral2018JPCL} or omittance of couplings between different PESs\cite{Chen2018JPCL} in ML driven photodynamics. Hence long time photodynamics are still lacking and the possibility to additionally represent the afore-mentioned nonadiabatic derivative couplings between PESs fundamental to model photodynamics has not been demonstrated yet.
Here we overcome all these different bottlenecks using deep neural networks (NNs) and achieve the simulation of photodynamics for long time scales. We expand on the idea of using ML to obtain potentials for electronic excited states, as well as arbitrary couplings within a framework that combines ML with trajectory surface hopping molecular dynamics (Fig. \ref{fig:1}). Our ML approach is fully capable of describing all necessary properties for executing nonadiabatic excited-state molecular dynamics on the order of nanoseconds. These properties include electronic energies, gradients, spin-orbit couplings, nonadiabatic couplings, and dipole moments of molecules. Additionally, the underlying potentials and couplings can be used to optimize critical points of the configurational space, such as potential minima or crossing points, which are critical for interpreting photochemical mechanisms. 

\section{Theoretical Background}

 Nonadiabatic excited-state molecular dynamics are carried out using the surface hopping including arbitrary couplings (SHARC) method,\cite{Richter2011JCTC} which is an extension of the fewest switches surface hopping method of Tully.\cite{Tully1990JCP} Within surface hopping, the nuclei are propagated according to the classical equations of motion and the electrons are treated quantum mechanically via interfaces to external electronic structure program packages. The electronic structure calculations are carried out on-the-fly at the nuclear geometries visited by the classical trajectories. The probability of a molecular system occupying a specific electronic state and population transfer between the different electronic states -- in the form of stochastic, instantaneous hops from one electronic state to another -- are dependent on the couplings between them. 

\subsection{Surface hopping molecular dynamics with deep NNs}
For surface hopping simulations with NNs, the idea of retrieving electronic properties from an external source stays the same, but instead of a quantum chemical calculation, NNs are used to predict energies, gradients, couplings and dipole moments. The relationships between the nuclear coordinates and the corresponding electronic properties are learned from a training set, in which each data point is one set of nuclear coordinates and its associated set of quantities computed with a reference method. In order to make the procedure usable, the processes for generating NNs potentials and their use in photodynamics simulations have been automated in a development version of the program suite SHARC.\cite{Richter2011JCTC,Mai2018WCMS,sharc-md2} 

\subsection{Training set generation and adaptive sampling for excited states}
The combination of quantum chemistry with ML requires a cost-effective generation of a training set that, while it samples the conformational space of a molecular system comprehensively, is small enough to keep demanding quantum chemical reference calculations feasible.\cite{Behler2017ACIE} With this in mind, we employ an initial training set based on normal mode scans and then switch to an adaptive sampling scheme\cite{Behler2015IJQC,Botu2015IJQC,Li2015PRL,Gastegger2017CS,Smith2018JCP} that automatically identifies untrustworthy regions not covered by the initial training set. The adaptive sampling procedure employs excited-state dynamics simulations using two or more NNs that are independently trained from the same training set. At every time step, the root mean squared error (RMSE) between the predictions of the different NNs of each property is compared to a predefined threshold. A separate threshold is set for each property (initially based on the validation error of the respective NN). Whenever any one of the thresholds is exceeded, i.e., the different NNs make very different predictions, the corresponding geometry is assumed to lie in a conformational region with too few training points, even if the rest of the properties is predicted reliably.
It is then necessary to expand the training set by computing the quantum chemistry data for this geometry. Along a dynamics run, the threshold for the error between predictions made by the NNs is adapted by multiplication with a factor of 0.95 until the conformational space is sampled sufficiently to make accurate predictions without any additional reference calculations. 

An ensemble of two NNs is used not only during the initial adaptive sampling period, but also for production dynamics simulations in order to check the accuracy of our NN predictions and to discover undersampled regions of conformational space. After 10 ps, the threshold for the RMSE between NN forecasts is not reduced anymore but kept at the previous value when a new data point is added to the training set and NNs are retrained. More details on criteria for the thresholds and iterations are discussed in the ESI\dag{}.

\subsection{Multi-layer feed-forward NNs}
For the sake of making predictions of quantum chemical properties of molecules, multi-layer feed forward NNs are applied.\cite{Behler2015IJQC} For training of NNs, we use as input the matrix of inverse distances in order to achieve translational and rotational invariance in the relations established between the predicted properties and the nuclear coordinates. For prediction we use two similarly accurate NNs, with their optimal-network-architecture identified by random grid search\cite{Goodfellow2016} of (hyper)parameters. Additional information on network parameters and specifications can be found in Table S1 and section S1 in the ESI\dag{} along with NN convergence during training in Fig. S1. We assessed the quality of the used NNs by comparing to different ML models and NNs using a different molecular descriptor on an additionally generated test set, see section S1.3 in the ESI\dag{}. Different ML models or descriptors do not lead to a considerable improvement of the accuracy. As a different ML model we choose support vector machine for regression and linear regression as a baseline model, whereas our NN approaches outperform  these regression models. Furthermore, the performance of our NNs is presented in Table S5 for each electronic state, separately. In this context, it is shown how the tendency towards smooth interpolation of the ML models can even correct for discontinuities present in the \textit{QC1} method (see Fig. S2), which demonstrates the utility of our approach.

Quantum chemical properties that were learned with NNs are energies, gradients, permanent as well as transition dipole moments, and NACs. Other quantities like spin-orbit couplings can also be trained (see analytical model in the ESI\dag{}). Although the (transition) dipole moments are not needed for the present dynamics simulation, calculating them on-the-fly enables the computation of pump-probe schemes, static-field interactions, or time-resolved spectra, see for example Refs. \cite{Marquetand2004JCP, Bonafe2018JPCL}. While energies are directly used for training purposes in a single NN, forces are predicted as analytical derivatives of the NNs,\cite{Gastegger2015JCTC} ensuring energy conservation.\cite{Gastegger2017CS,Hu2018JPCL,Christensen2019JCP} Similarly, permanent dipole moments are directly used in the training. However, couplings (as well as transition dipole moments) need to be pre-processed as they are computed from the wave functions of two different electronic states and therefore depend on the relative phases of these two wave functions. Phase inconsistencies need to be eliminated in order to avoid ill-behaved photodynamics,\cite{Akimov2018JPCL} as it is described in the following subsection. 

\subsection{Phase correction}
\begin{figure}[h]
\centering
  \includegraphics[height=8.5cm]{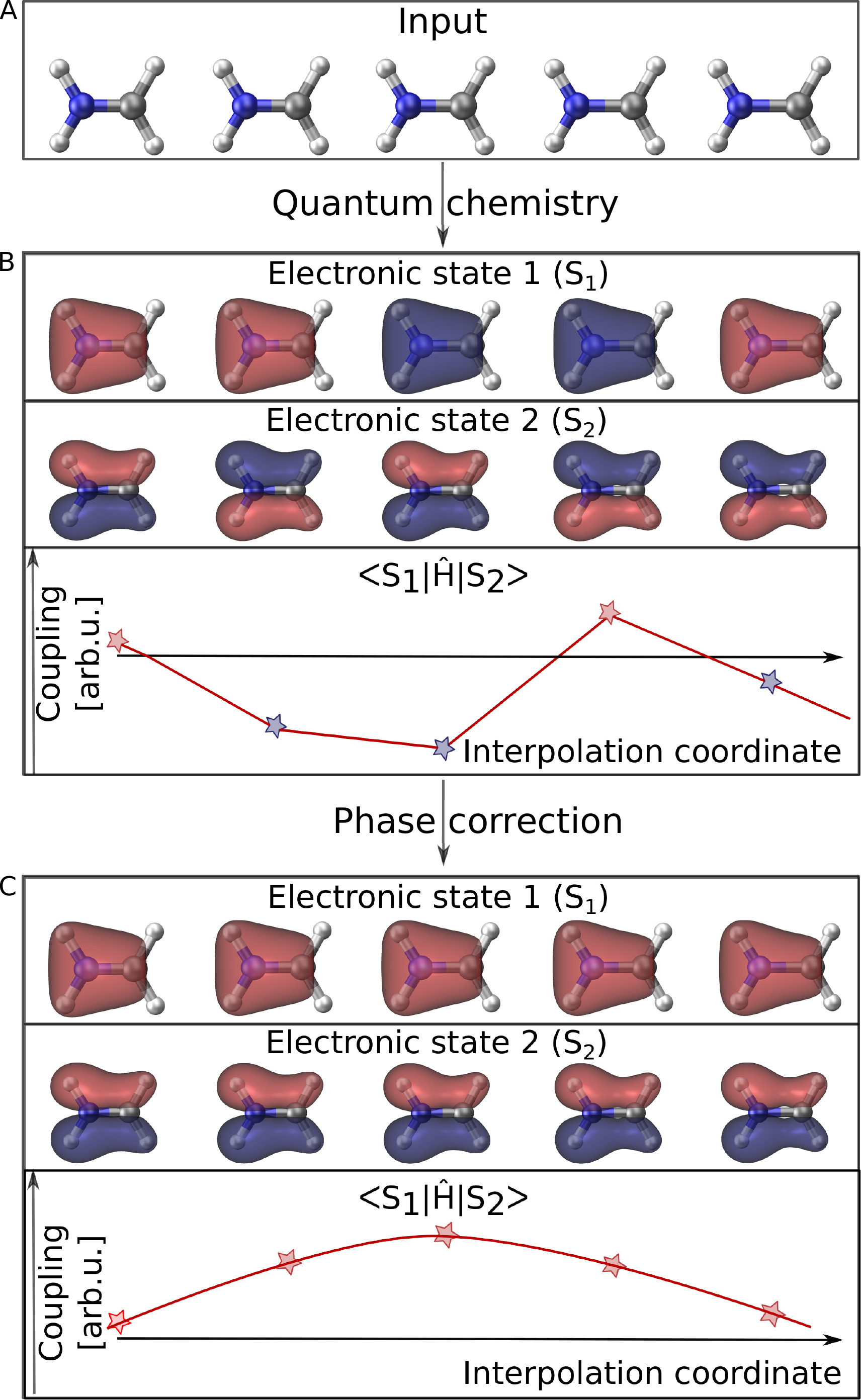}
  \caption{Molecular orbitals representing two different electronic states of the methylenimmonium cation, CH$_2$NH$_2^+$. Panel A shows molecular geometries (with slightly different bond lengths) that are given as an input to a quantum chemistry program. The results for properties corresponding to off-diagonal matrix elements of the Hamiltonian are shown in the panel B. Random signs are obtained due to random assignments of the phases of the involved wave functions. As can be seen in panel C, these random switches can be removed with phase correction and smooth relations between a molecular geometry and any property can be found.}
  \label{fig:2}
\end{figure}

Electronic wave functions computed with quantum chemistry programs are usually obtained as the eigenfunctions of the electronic Hamiltonian. However, this requirement does not uniquely define an electronic wave function because multiplying it by a phase factor still returns a valid eigenfunction. Thus, in practice two wave functions computed for two very similar geometries might randomly differ by their phase factor. This problem is best visualized using molecular orbitals, see Fig. \ref{fig:2}. For different single point calculations along an interpolation coordinate (Fig. \ref{fig:2}A), orbitals can arbitrarily switch their sign (illustrated by their color in Fig. \ref{fig:2}B) and so does the complete electronic wave function. As energies are obtained from diagonal elements in matrix notation, the electronic wave function of the general form $\langle \Psi_i \mid \hat{O}\mid\Psi_i\rangle$, enters twice and any phase is squared, thus canceling out. However, off-diagonal elements, $\langle \Psi_i \mid \hat{O}\mid\Psi_j\rangle$, such as couplings involve the wave functions of two different electronic states and different phases do not necessarily cancel out. The example of Fig. \ref{fig:2}B shows how the curves of such off-diagonal properties can be discontinuous, impeding correct learning behavior in the NN. It is thus mandatory to track the phases of all wave functions from one reference geometry to every other data point in the training set and apply a phase-correction algorithm that provides smooth curves (Fig. \ref{fig:2}C). In this way, a virtual global phase convention is applied to all data points within the training set, with the only aim of ensuring correct NN training. 

Such a global phase convention is not mathematically possible for general poly-atomic molecules due to the existence of the so-called Berry (or geometric) phase.\cite{Ryabinkin2017ACR} Due to the latter, the phase depends on the path between a given geometry and the reference point.\cite{Matsika2011ARPC}
Still the above phase correction is advantageous because it removes phase jumps from almost all parts of configuration space. This is critically necessary to make the data learnable. Only the non-removable phase jumps from the Berry phase remain, but occupy a small volume of configuration space.
Hence, our phase correction is assumed to leave the dynamics mostly unaffected. For instance, successful surface hopping algorithms without phase tracking, such as the Zhu-Nakamura theory,\cite{Oloyede2006JCP,Ishida2017IRPC} exist and substantiate the validity of this approximation. In the case of the Zhu-Nakamura theory, dynamics are comparable to conventional surface-hopping molecular dynamics simulations propagated from NAC vectors.\cite{Zhu2002JCP,Ishida2009JPCA,Gao2012JCP,Yu2014PCCP,Ishida2017IRPC} Note that the approximated phase correction for generation of the training set above cannot be circumvented by learning the absolute value of couplings since the relative sign between nonadiabatic coupling vectors of each atom in x, y and z direction should be retained. 

In order to make off-diagonal elements learnable for ML models, phases are tracked by computing wave function overlaps between adjacent molecular geometries.\cite{Mai2015IJQC,Plasser2016JCTC,Akimov2018JPCL} If the geometries are close enough, the overlaps will be sufficiently large and contain values close to +1 or -1, allowing a detection of phase changes. In cases where molecular geometries are too far apart, the overlap will generally be close to zero, offering no information about a phase change. In this case, we resort to interpolation between the two molecular geometries and iterative computation of wave function overlaps. In principle, the interpolation can be carried out between the new geometry and any geometry already inside the training set as long as the wavefunction of this previous geometry is stored. Storing the wavefunctions for at least a few geometries and identifying the most suitable one for interpolation via root mean square deviations of the geometry should be considered for larger and more flexible molecules.

Especially for large molecules, where many states lie close in energy, so-called ''intruder states'' might become problematic. Such states are excluded at the reference geometry, but are included at another geometry due to an energy change, thus leading to small overlaps for the phase tracking algorithm. In such situations, different possibilities for adapting the phase correction algorithm should be considered.For instance, additional electronic states could be computed with QC. Those should not be included in the training data, but only used to continuously track the phase of all relevant states. This process then still stays affordable, since the additional states do not require a computation of gradients or couplings and do not have to be considered further. Additional details on the phase correction algorithm are given in section S2 in the ESI.\dag

\section{Computational details}

The photodynamics simulations have been carried out with a development version of the program suite SHARC.\cite{Mai2018WCMS,sharc-md2} Besides the newly developed modules for NN training and prediction, this development version also employs the pySHARC Python wrapper for the SHARC dynamics driver. This wrapper enables communication between the driver and the NN code without any file I/O and thus reduces the runtime of the program substantially.

The reference quantum chemical computations were carried out with COLUMBUS\cite{Lischka2001PCCP} using the accurate multi-reference configuration interaction method including single and double excitations and a double-zeta basis set (abbreviated as MR-CISD/aug-cc-pVDZ and in the following sections labelled as \textit{QC1}). 
For comparison, we carried out quantum chemical computations with another basis set, 6-31++G**, using the same MR-CISD method (abbreviated \textit{QC2} in the following sections). NNs were implemented in Python using the numpy\cite{Walt2011CSE} and theano\cite{TDT2016a} packages. They were trained on energies, forces, dipole moments and nonadiabatic couplings, obtained with the \textit{QC1} method using the adaptive sampling scheme described above, resulting in about 4000 data points (Mean absolute error (MAE) energies among all states: 0.032 eV $\hat{=}$ 0.73 kcal/mol; MAE forces among all states: 0.51 eV/{\AA} $\hat{=}$ 11.9 kcal/mol/{\AA}, see also Table S2, S4 and S5 in the ESI as well as Fig. S2 for analysis of different states\dag). With each method, \textit{QC1}, \textit{QC2}, and NNs trained on \textit{QC1}, we simulated the dynamics of the methylenimmonium cation after excitation from the electronic ground state (S$_0$) to the second excited electronic state (S$_2$) during 100 fs using a time step of 0.5 fs. 

Optimizations of minima were carried out with the SHARC tools that utilize an external optimizer of ORCA,\cite{Neese2012WCMS} where the computed energies and gradients\cite{Levine2007JPCB,Bearpark1994CPL} from the NNs were fed in or those from COLUMBUS for comparison.

\section{Results and Discussion}
First, a one-dimensional model was employed to test our deep learning molecular dynamics approach (see Fig. S3 in section S3 in the ESI\dag). In the following, the performance of the method is demonstrated by simulating the full dimensional photodynamics of the methylenimmonium cation, CH$_2$NH$_2^+$ - the simplest member of the protonated Schiff bases. Methylenimmonium has been reported to undergo ultrafast switches between different electronic states after excitation with light.\cite{Barbatti2006MP} A larger member of this family is retinal, which is fundamental for vision\cite{Herbst2002S} but the methylenimmonium cation is an ideal testbed to demonstrate the applicability of NNs in photodynamics, because it is small enough to perform accurate reference photodynamics simulations for short time scales for comparison.

\subsection{Nanosecond molecular dynamics simulation}

\begin{figure}[h]
\centering
  \includegraphics[height=4cm]{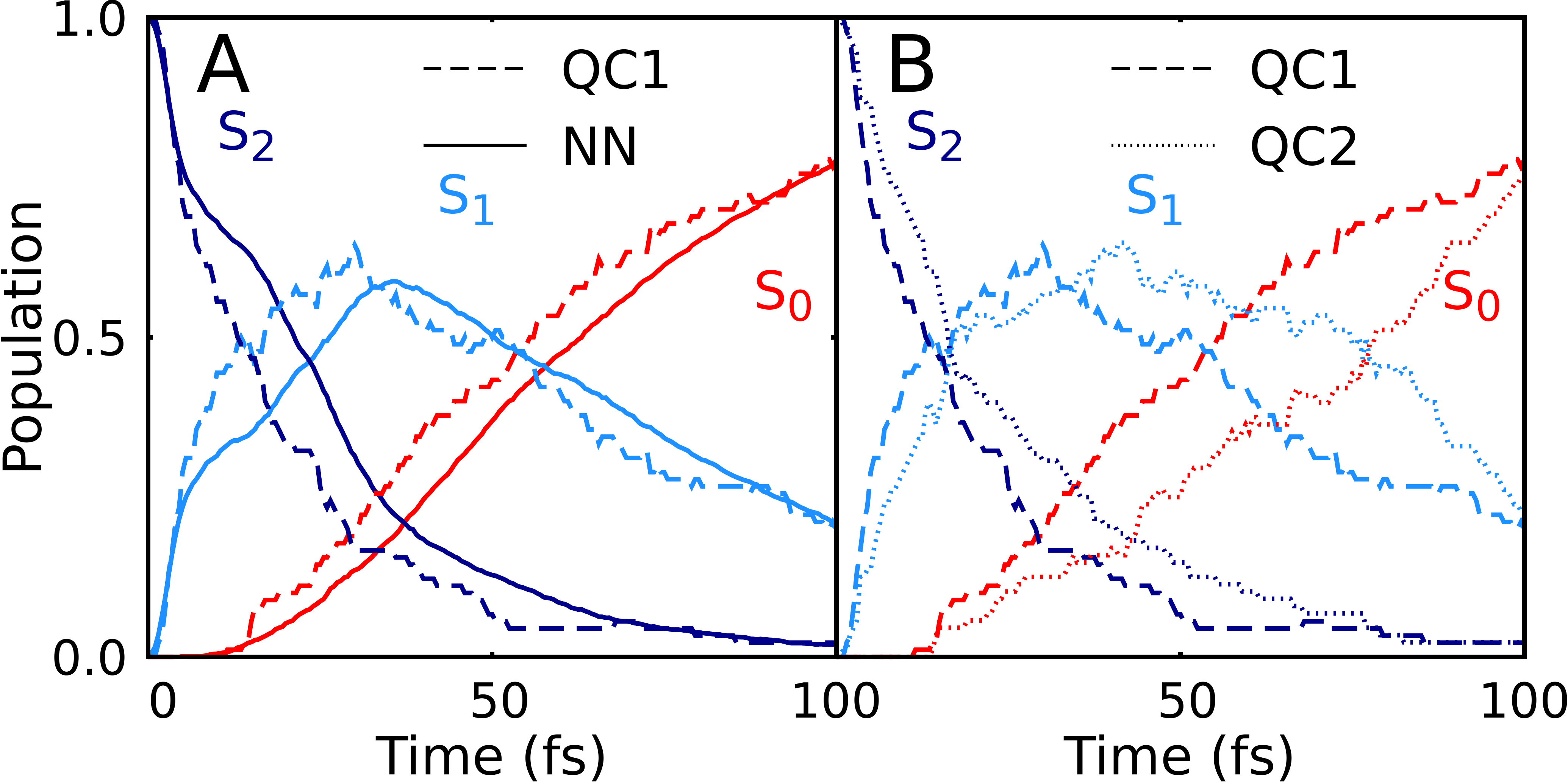}
  \caption{Population dynamics of CH$_2$NH$_2^+$ based on deep neural networks and traditional quantum chemistry:  Comparison between results obtained from (A) \textit{QC1} (90 trajectories) and neural networks (NN, 3846 trajectories) and (B) \textit{QC1} (90 trajectories) and \textit{QC2} (88 trajectories). 
  For completeness, the populations from 90 trajectories propagated with NNs are given in Fig. S5A in the ESI\dag ~along with geometrical analysis along the trajectories in Fig. S5B .}
  \label{fig:3}
\end{figure}

Our NNs were trained on data obtained with the \textit{QC1} method (see details on active space in section S4 and Fig. S4 in the ESI\dag). Independently with the \textit{QC1} method and with NNs, we simulated the dynamics of the methylenimmonium cation after excitation to the second excited singlet state, S$_2$. As can be seen from Fig. \ref{fig:3}A, fast population transfer from the S$_2$ state to the first excited singlet state, S$_1$, and back to the ground state, S$_0$, takes place. The population dynamics obtained with the NN potentials and that obtained using the \textit{QC1} method agree very well. These results are also in good agreement with literature.\cite{Barbatti2006MP} Both methods describe the deactivation to the ground state, S$_0$, through the correct conical intersections, as will be discussed in the next subsection. 

One first advantage of the NN driven dynamics simulations is that due to its very low computational cost, a much larger number of trajectories (3846) was simulated than what is typically possible with standard quantum chemistry (90). This enlarged statistics provides smooth population curves for the NNs simulations (a comparison of the curves with identical number of trajectories for NNs and \textit{QC1} can be found in Fig. S5A in the ESI\dag{} along with analysis on energy conservation in Table S11.). 

In order to estimate the magnitude of the error obtained with the NNs, we carried out a second ab-initio molecular dynamics study with an additional, very similar, quantum chemistry method where only the double-zeta basis set is changed;  from aug-cc-pVDZ to 6-31++G**.
As Fig. \ref{fig:3}B shows, the differences between both levels of theory are of the same order of magnitude as those encountered between NNs and quantum chemistry, indicating that the agreement between the methods is very good. The MAE in population between \textit{QC1} and NNs is 0.057 and between \textit{QC1} and \textit{QC2} 0.099. Time constants derived from dynamics with each method also agree well. The time constant from S$_2$ to S$_1$ is 18.3 fs according to the \textit{QC1} method, which is comparable to the \textit{QC2} method with 25.0 fs and to NNs driven dynamics with 25.2 fs. The time constant obtained for transitions from S$_1$ to S$_0$ is 51.0 fs for the \textit{QC1} method, which is very similar to the value obtained with NNs (52.6 fs), whereas the \textit{QC2} method yields a time constant of 73.2 fs.

\begin{figure}[h]
\centering
  \includegraphics[height=3.3cm]{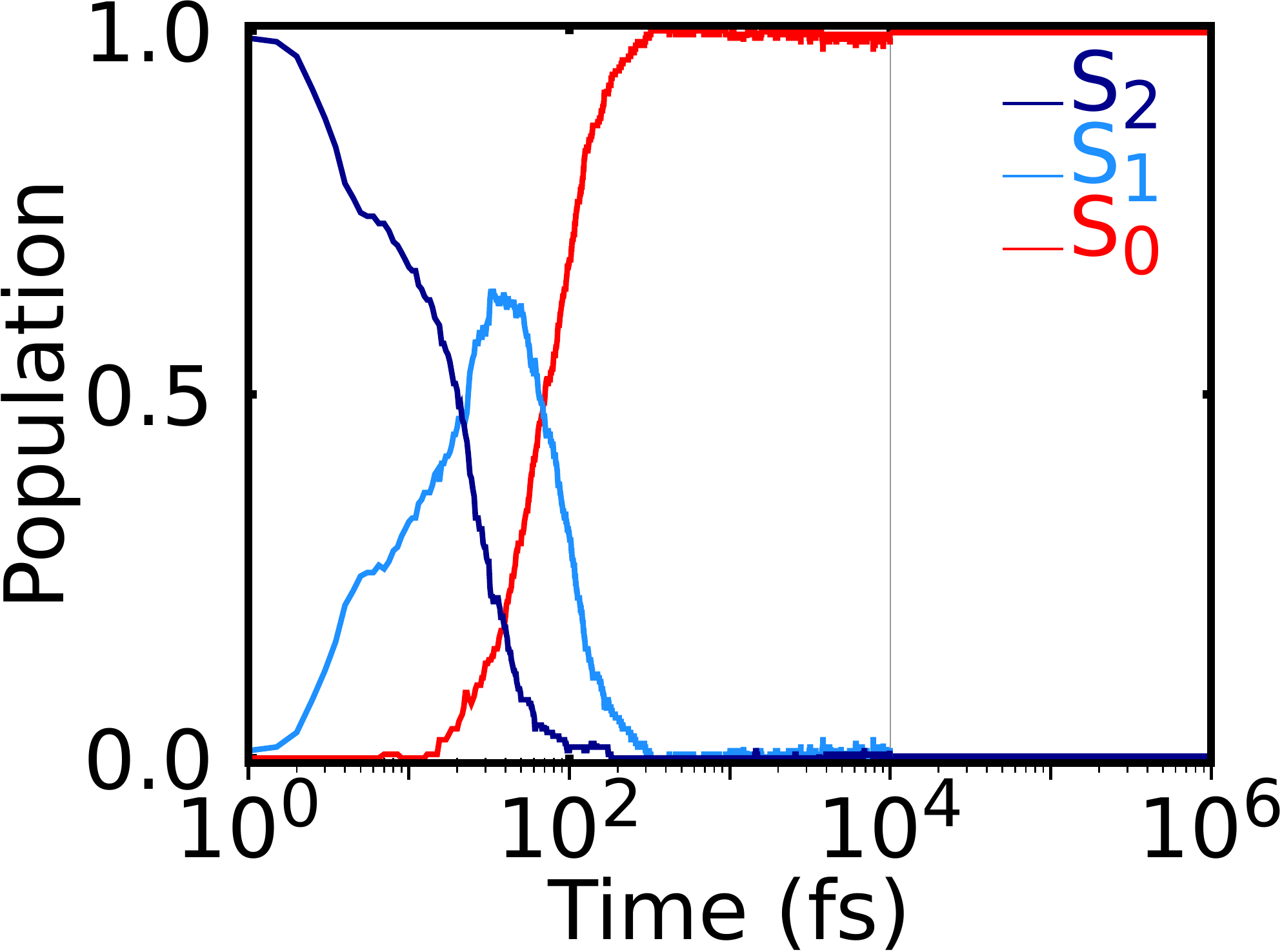}
  \caption{Nonadiabatic molecular dynamics simulations using deep NNs for one nanosecond. After excitation to the S$_2$ state, ultrafast internal conversion to the S$_1$ state takes place, followed by recovery of the S$_0$ state within 300 fs. Until 10 ps, an ensemble of 200 trajectories is analyzed, followed by the population averaged from 2 trajectories.}
  \label{fig:4}
\end{figure}

After nonadiabatic dynamics using deep NNs has been validated for short time scales, we show the major advantage of the method, i.e. that it is able to overcome the problem of limited simulation time and predict long excited-state dynamics. Fig. \ref{fig:4} shows the population dynamics of the methylenimmonium cation on a logarithmic scale up to 1 nanosecond (ns), i.e., 10$^4$ times longer than they were simulated with our quantum chemical reference method. 
Up to 10 ps, we simulated an ensemble of 200 trajectories with 2 NNs using the adaptive sampling scheme described above in order to correctly predict events not yet learned by the NNs. After that, 2 trajectories are propagated up to 1 ns for demonstration purposes using 2 NNs. The populations are thus averaged over 200 trajectories up to 10 ps and over 2 trajectories from 10 ps on up to 1 ns, respectively. As can be seen, the molecule relaxes to the ground state after around 300 fs. Due to remaining kinetic energy a few hops between different states are recorded and can be regarded as noise.  

The propagation of CH$_2$NH$_2^+$ for 10 ps can be executed in less than 6 hours on one core, which is 300 times faster than the calculation with the quantum chemical reference method. The propagation of 1 ns took 59 days with deep NNs, whereas an estimated \textasciitilde 19 years of computation would have been required with the quantum chemical reference.

\subsection{Conical intersections obtained from NN}

Since the NNs can provide energies, gradients, and couplings, they can also be used to optimize important points of the PES, like state minima or conical intersections. The identification of conical intersections is the target of many quantum chemical studies as they are commonly deemed as the most probable geometries for radiationless transitions between electronic states of the same spin multiplicity. Due to their special topology with discontinuous first derivatives, the surrounding of a conical intersection poses serious challenges to the NN training.\cite{Domcke2004} As the photodynamics critically depends on a correct representation of these surroundings, here we perform some tests to validate their accuracy.

\begin{figure}[h]
\centering
  \includegraphics[height=6cm]{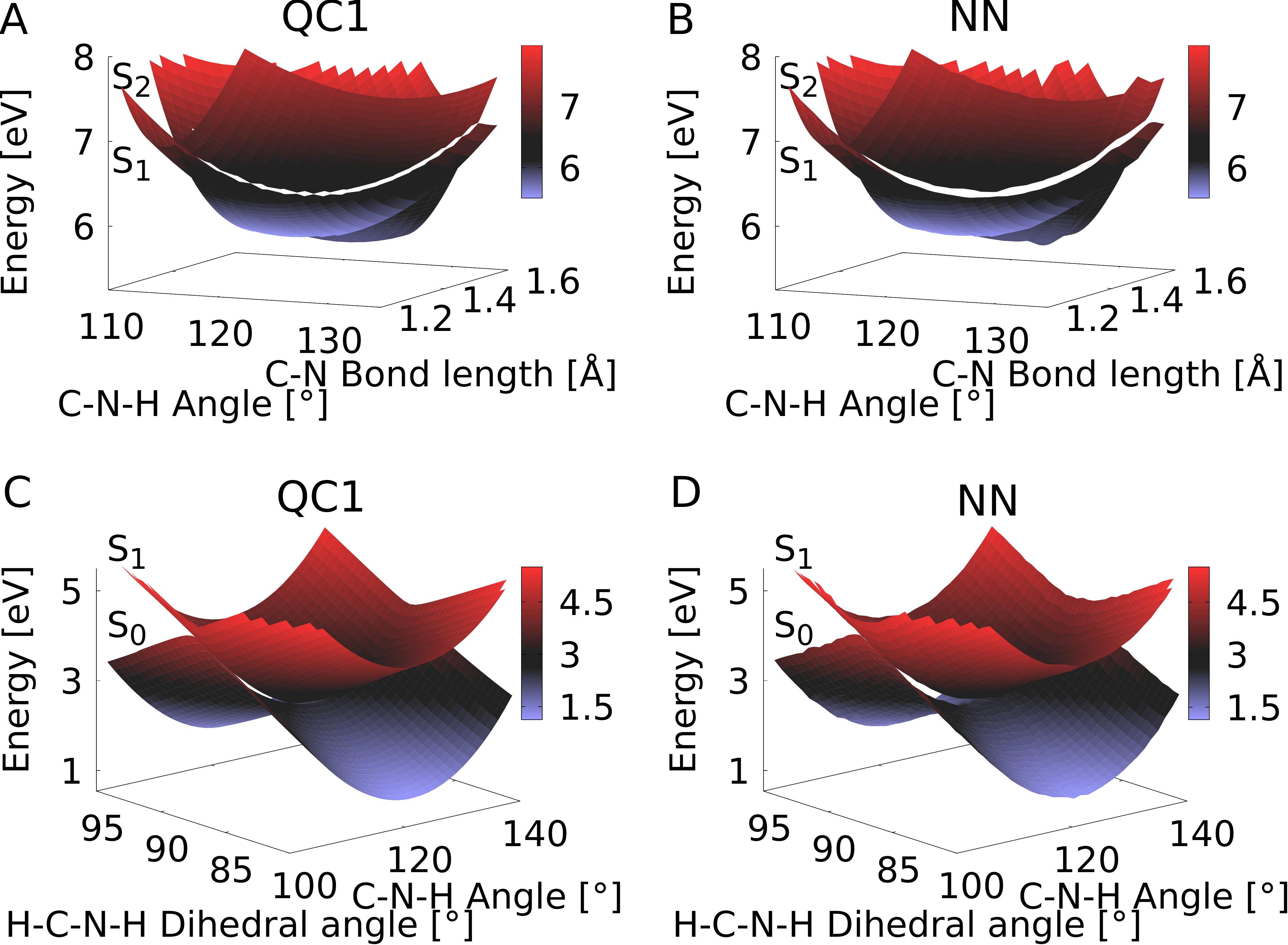}
  \caption{Potential energy scans around the minimum energy conical intersections obtained with \textit{QC1} of the S$_2$ and S$_1$ state (5A-B) and S$_1$ and S$_0$ state (5C-D). Panels A  and C show the PESs calculated with \textit{QC1}, whereas panel B and D illustrate NN potentials. See caption of Fig. S7 in the ESI\dag ~for clarification of the dihedral angle.}
  \label{fig:5}
\end{figure}

To this aim we optimize two minimum energy conical intersections in CH$_2$NH$_2^+$, one between the S$_2$ state and the S$_1$ state and another one between the S$_1$ state and the S$_0$ state. We use the \textit{QC1} method and NNs to perform potential energy scans around the minimum energy conical intersections optimized at the \textit{QC1} level of theory. As can be seen from Fig. 5A-D, typical curved seams of conical intersections between the S$_2$ and S$_1$ state (Fig. 5A (\textit{QC1}) and 5B (NN)) and the S$_1$ and S$_0$ state (Fig. 5C (\textit{QC1}) and 5D (NN)) are obtained around the minimum energy conical intersections.\cite{Yarkony2005JCP} The NNs get the shape of this seam correct with slightly larger energy gaps between the crossing surfaces due to the fact that NN potentials need to be differentiable at any point. Analysis of 408 (for the S$_1$/S$_0$ CI) and 302 (for the S$_2$/S$_1$ CI) configurations around the minimum energy conical intersections -- identified by an energy gap smaller than 0.8 eV according to the \textit{QC1} method -- showed that on average, the gaps are overestimated by 0.068 eV for S$_1$/S$_0$ and by 0.014 eV for S$_2$/S$_1$ by our NNs. As can be seen from Fig.~\ref{fig:5}, the potentials around the S$_1$/S$_0$ CI are flatter than the potentials around the S$_2$/S$_1$ CI, indicating that hopping geometries are closer to the CI in the latter case and that the molecules can also hop farther from the CI in the former case.

\begin{figure}[h]
\centering
  \includegraphics[height=8.5cm]{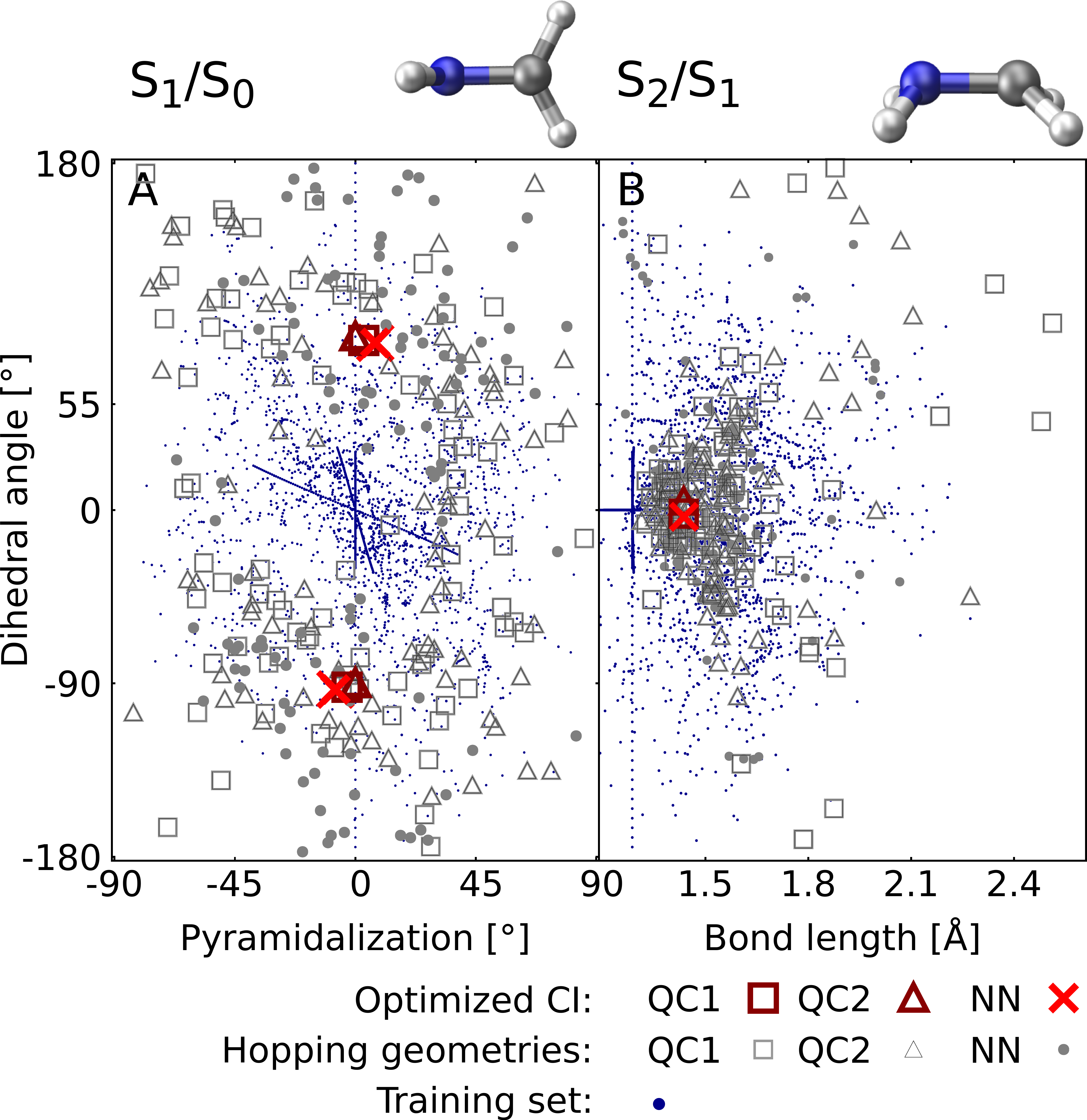}
  \caption{Scatter plots showing the distribution of hopping geometries obtained with \textit{QC1}, \textit{QC2}, and NN as well as optimized S1/S0 (A) and S2/S1 (B) minimum energy conical intersections (CI) along with the geometries that make up the training set with 4000 data points. The actual geometry is depicted on top (geometrical parameters are given in Fig. S7B). A zoom of the regions near the optimized points is shown in Fig. S7A in the ESI\dag ~together with a definition of the dihedral and pyramidalization angles.}
  \label{fig:6}
\end{figure}

Fig. \ref{fig:6} shows the scatter plots of the optimized geometries of the minimum energy conical intersections projected along two important coordinates together with the hopping geometries and the geometries contained in the training set. As can be seen, the hopping geometries between the S$_2$ and S$_1$ state are mainly located close to the optimized geometry of the minimum energy conical intersection, while the hopping geometries in case of the S$_1$/S$_0$ crossing are more widely distributed around the optimized geometry. As a consequence, the S$_2$/S$_1$ crossing is sampled more comprehensively, since more trajectories pass by near the minimum energy conical intersection. This observation also explains the larger NN energy gap obtained for the second crossing, the S$_1$/S$_0$ CI, in Fig. \ref{fig:5}.

The optimizations of the minimum energy conical intersections were independently performed with the trained NN, as well as with the \textit{QC1} and \textit{QC2} methods for comparison. The optimized molecular geometries (shown in Fig. S6 along with Cartesian coordinates in the ESI\dag) agree well. As can be seen, the driving force for the transition from the S$_2$ state to the S$_1$ state is an elongation of the C-N bond in combination with a bipyramidalization. The torsion of the molecule further leads to internal conversion to the ground state, S$_0$. 
Additionally, each method results in a comparable distribution of hopping geometries around the optimized points, which in practice is of uttermost importance\cite{Hudock2007JPCA} for describing the population transfer in the simulations correctly. 
There are very few NN hopping geometries at either large pyramidalization angles (S$_1$/S$_0$ CI) or long C-N bonds (S$_2$/S$_1$ CI), compared to the QC trajectories. This finding correlates with the distribution of training set geometries, which are also absent in these regions of the PES, see the blue circles in Fig.~\ref{fig:6}. Configurations obtained via sampling of normal modes are clearly visible by a dense alignment of data points. However, the configurations obtained via adaptive sampling are mostly centered in the middle of the plot for the S$_1$/S$_0$ CI and close to the optimized CI for the S$_2$/S$_1$ crossing, explaining the smaller distribution of NN hopping geometries. Further analysis showed that geometries at large bond lengths are approximately 4 eV higher in energy than the geometries close to the optimized minimum energy conical intersection in case of the S$_2$/S$_1$ crossing. Therefore, trajectories carried out during adaptive sampling probably did not visit those regions of PES. In case of the S$_1$/S$_0$ crossing, this effect is less pronounced and the geometries with a large pyramidalization angle are approximately 1-1.5 eV larger in energy than the configurations close to the optimized CI, indicating again the much flatter potential.

\section{Conclusions}
We show that deep NNs are able to accelerate nonadiabatic excited-state molecular dynamics simulations by orders of magnitude, thus overcoming the constraints of limited time scales and limited statistics. Our approach offers an automatic learning procedure by implementation of adaptive sampling for excited states, which opens new avenues for studying the photodynamics of complex systems on long time scales relevant for chemistry, biology, medicine, and material design, for which the PESs cannot be explored in advance with conventional ab-initio techniques. Offering access to the precision of high-level quantum chemistry methods at only a fraction of the original computational cost, we expect this setup to become a powerful tool in several research fields. 

\section*{Conflicts of interest}
There are no conflicts of interest to declare.
\section*{Acknowledgements}
We thank Anna B\"{a}ck for her contributions to reference calculations. This work was financially supported by the uni:docs program of the University of Vienna (J.W.), the European Union Horizon 2020 research and innovation program under the Maria Sklodowska-Curie grant agreement NO 792572 (M.G.), the ITN-EJD 642294 (M.F.S.J.M.), and the Austrian Science Fund (I 2883). The computational results presented have been achieved in part using the Vienna Scientific Cluster (VSC).





\providecommand*{\mcitethebibliography}{\thebibliography}
\csname @ifundefined\endcsname{endmcitethebibliography}
{\let\endmcitethebibliography\endthebibliography}{}

\clearpage
\renewcommand{\thefigure}{S\arabic{figure}} 
\renewcommand{\thetable}{S\arabic{table}} 
\renewcommand{\thepage}{\arabic{page}} 
\renewcommand{\thesection}{S\arabic{section}}
\setcounter{figure}{0}
\setcounter{table}{0}
\setcounter{section}{0}
{\onecolumn
\noindent\normalsize{Electronic Supplementary Material (ESI) \\
\vspace{0.5cm} \\
\renewcommand{\baselinestretch}{1.5}
\noindent\LARGE{\textbf{Supporting Information: Machine learning enables long time scale molecular photodynamics simulations}} \\
\noindent\large{Julia Westermayr,\textit{$^{a}$} Michael Gastegger,\textit{$^{b}$} Maximilian F. S. J. Menger,\textit{$^{a,c}$} Sebastian Mai,\textit{$^{a}$} Leticia Gonz\'{a}lez\textit{$^{a}$} and Philipp Marquetand$^{\ast}$\textit{$^{a}$}} \\
}

\renewcommand*\rmdefault{bch}\normalfont\upshape
\renewcommand{\baselinestretch}{1.5}
\renewcommand{\arraystretch}{1.5}

\rmfamily
\section*{}
\vspace{-1cm}


\footnotetext{\textit{$^{a}$~Institute of Theoretical Chemistry, Faculty of Chemistry, University of Vienna, 1090 Vienna, Austria. }}
\footnotetext{\textit{$^{b}$~Machine Learning Group, Technical University of Berlin, 10587 Berlin, Germany. }}
\footnotetext{\textit{$^{c}$~Dipartimento di Chimica e Chimica Industriale, University of Pisa, Via G. Moruzzi 13, 56124 Pisa, Italy. }}
\footnotetext{\textit{$^{\ast}$~ Corresponding author. E-mail: philipp.marquetand@univie.ac.at }}




\section{Methods}

\subsection{Neural networks (NNs)}

Multi-layer feed forward neural networks (NNs) have been implemented in python using the numpy\cite{Walt2011CSE} and theano\cite{TDT2016a} packages.

To find (hyper)parameters of the optimal-network-architecture NNs to best fit the relation between a molecular geometry and its corresponding excited-state properties, we have automated a random grid search\cite{Goodfellow2016} and adapted the learning rate, one of the most critical parameters during training.\cite{Glorot2010} The optimization of several parameters of NNs was done for the initial training set consisting of 100 data points for the analytical model and 992 data points for CH$_2$NH$_2^+$. After the adaptive sampling, the training set increased to twice its initial size. Each nuclear configuration was given to the NNs as the matrix of inverted distances. For optimal performance, we tested several aspects of the NN architecture, like the type of non-linear basis function (hyperbolic tangent and shifted softplus function, $ln(0.5 e^{x}+0.5)$) used, the number of neurons, the number of hidden layers, as well as other parameters, such as the learning rate, $lr$ in the following equations, the L2 regularization rate, the number of epochs, the batch size, and a constant factor, $\eta$ in equation \ref{eq:2}, that regulates the influence of forces in the update step. Emphasis was put on optimization of the L2 regularization rate with respect to a given batch size and the learning rate. 

For each quantity that has to be predicted, a separate NN was used, except for the gradients. The latter enter directly into the NNs dedicated to the potential energy prediction. The NNs for the analytical model consisted of one or two hidden layers, whereas deep NNs with 6 hidden layers were applied for training and predictions of properties of CH$_2$NH$_2^+$. We chose the NNs with respect to their error on the validation set as well as the form of their loss function. 
For the initial training set, we always sampled two different variables of the NN architectures. The learning rate was first scanned during a couple of iterations from values of 10$^{-7}$ to 10$^{0}$ along with the L2 regularization rate (values between 10$^{-4}$ and 10$^{-14}$) using a fixed number of neurons (50) and hidden layers (4). After getting an idea about the magnitude of the L2 regularization rate, the decay factor as well as the step number for annealing the learning rate were evaluated. The decay factor was sampled from 0 to 1, whereas values between 0.9 and 1.0 seemed to give best NN training. The number of epochs, after which the learning rate should be annealed, was sampled from 1 to 1000 with values between 1 and 100 leading to best results. After this, the type of basis function was analyzed. The hyperbolic tangent turned out to give fair results for nonadiabatic couplings (NACs) and dipole moments, whereas the shifted softplus function was slightly better for energies and gradients. We further tried batch sizes of 1, 2, 5, 10, 32, 64, 100, 128, 256, 500, and 992 for the initial training set (batch sizes of 2000 and 4000 were also tested for the larger and final training sets) along with the learning rate. A batch size of 500 turned out to lead to good training performance. The number of hidden layers was evaluated from 1 hidden layer up to 8 hidden layers along with the number of neurons from 10 to 100. For energies, gradients and dipole moments the number of neurons was set to 50 and for networks trained on NACs a slightly larger number of neurons was chosen. After assessing those parameters, the learning rate as well as the L2 regularization rate were sampled again. Now, a narrower space of those parameters was sampled relying upon the previously defined values. The influence of the decay factor and update step for annealing the learning rate was further reassessed. 4 Hidden layers were used for training set sizes up to around 2000 data points, afterwards 6 hidden layers slightly improved the NN accuracy.

Each time new parameters were sampled, we trained around 20-100 networks depending on the space to cover for a given parameter. The large initially sampled space of (hyper)parameters was narrowed after the initial training set was expanded and assumed to be a good starting point for random grid search. There were several sets of parameters that led to very similar performance and the two with the lowest error on a validation and training set were chosen for carrying out dynamics simulations.

\subsection{NN training}
In general, training aims at fitting the weight parameters of a NN in order to minimize a cost function, which usually represents the mean-squared error (MSE) between predictions by NNs and reference data. For this task, we used Adam (adaptive moment estimation),\cite{Adam2014} a stochastic gradient descent optimization algorithm. In combination with Adam, an exponential learning rate decay was applied.\cite{Goodfellow2016} Therefore, a decay-factor, $f_{lr}$, with values between 0 and 1, as well as a step size were defined. The step size gives the number of epochs that have to be passed after $lr$ is adjusted according to the following equation:
\begin{equation}
lr = lr \cdot f_{lr}
\end{equation}		

Quantities, which can be related to a nuclear configuration and predicted by implemented NNs are energies as well as corresponding gradients, spin-orbit couplings, NACs, and dipole moments. For predictions of energies, it is favorable to include gradients in the minimization process.\cite{Gastegger2017CS} The cost function,
\begin{equation}
\label{eq:2}
C_{E,F} = \frac{1}{M} \sum_m^M (E_m^{NN}-E_m^{QC})^2+\frac{\eta}{M}\sum_m^M\frac{1}{3N_m}\sum_\alpha^{3N_m}(F_{m_\alpha}^{NN}-F_{m_\alpha}^{QC})^2
\end{equation}
thus depends on two terms, with m running over all molecules (total number M), and $\alpha$ running over all Cartesian coordinates of atoms (total number 3N). The first part of Equation \ref{eq:2} is the MSE between energies predicted by the NN, $E_m^{NN}$, and reference-data energies, $E_m^{QC}$, obtained from quantum chemical calculations. The second part of this equation represents the MSE of molecular forces, once derived from NNs, $F_{m_\alpha}^{NN}$, and once calculated with quantum chemistry, $F_{m_\alpha}^{QC}$. The constant $\eta$ regulates the influence of forces on the overall cost function and is set to 1 here. All different electronic states were treated within one NN, thus the errors entering the cost function are for all electronic states. 

The reference data set was always split into a training and a validation set in a random fashion using a ratio of 9:1. For training and prediction, inputs and outputs, $X$, were scaled with respect to the mean, $\mu$, and standard deviation, $\sigma$, of the training set according to equation \ref{eq:3}.
\begin{equation}
\label{eq:3}
    s = \frac{X-\mu}{\sigma}
\end{equation}

An early stopping mechanism was used to control overfitting. Additionally, the convergence of the loss functions of the training and validation set were checked manually, ensuring, e.g., that their order of magnitude is similar. The training set of the analytical model consisted of 100 equidistant data points containing values of x between 0.4 and 1.8 (with x being the one degree of freedom of the harmonic oscillators, see Table S7). Two initial training sets were generated for CH$_2$NH$_2^+$: one via sampling along each degree of freedom as well as a torsional mode and another one by executing dynamics simulations with SHARC but starting each trajectory from the same geometry, i.e., the equilibrium geometry, with sampled velocities according to a Wigner distribution.\cite{Wigner1932PR} The excitation window for these simulations was set to 7-11 eV. The training set sampled along normal modes and the torsional mode yielded a better initial fit of the different computed properties, thus the focus was set only on this set for CH$_2$NH$_2^+$. The initial data set was insufficient for excited-state dynamics simulations and was expanded via an adaptive selection scheme as described in reference.\cite{Gastegger2017CS} For this reason, two slightly different sets of NNs (NN1 and NN2, see Table \ref{tab:1} for the differing parameters and Table \ref{tab:MAE_molecule} for the mean absolute error (MAE) obtained with those NNs) were chosen to start dynamics simulations by replacing quantum chemical calculations with predictions made by NNs. The mean, $\overline{M}$ , of fitted properties, $\widetilde{M_J}$, of each NN, $J$,
\begin{equation}
    \overline{M}=\frac{1}{J}\sum_{j=1}^J\widetilde{M}_J
\end{equation}
is given to SHARC to propagate the nuclei. Further, the root mean-squared error (RMSE) between those predictions, $M_\sigma$, was calculated on-the-fly:
\begin{equation}
    M_\sigma = \sqrt{\frac{1}{J-1} \sum_j^J(\widetilde{M}_J-\overline{M})^2}.
\end{equation}
Inputs according to a region of the PES that has not been visited yet could be detected on-the-fly by setting a threshold for the RMSE of each quantity predicted by NNs. The threshold is set to 18.8 kcal/mol (0.03 H) for the energies in the beginning and adapted to smaller values during sampling similar to reference\cite{Gastegger2017CS} by multiplication with a factor of 0.95 up to 10 times. Whenever this threshold is exceeded, predictions by NNs are deemed untrustworthy and need to be recomputed with quantum chemistry as well as added to the training set. This is done for each property independently and as soon as one property is predicted unreliably, a new QC reference calculation is carried out. The NNs are retrained and previous weights are not used as a guess for training NNs. For the dipole moments, the threshold was set to 0.5 a.u. (atomic units). The latter was kept constant, since dipole moments do not directly enter into the dynamics in our current approach but are implemented for future research purposes.

For the RMSE of the NACs, the threshold is initially set to 0.25 a.u., reduced adaptively, and raised intermediately to 3.0 a.u. after a training set size of 3600 points is reached. This intermediate raise makes it possible to focus only on geometries close to conical intersections, where the NAC is verly large. These are necessary to describe the dynamics accurately. Consequently, the population dynamics converges towards the correct behavior, as can be seen in Fig. \ref{fig:adaptivesampling} for the last iterations of the sampling procedure. 

\begin{figure}
    \centering
    \includegraphics[height=10cm]{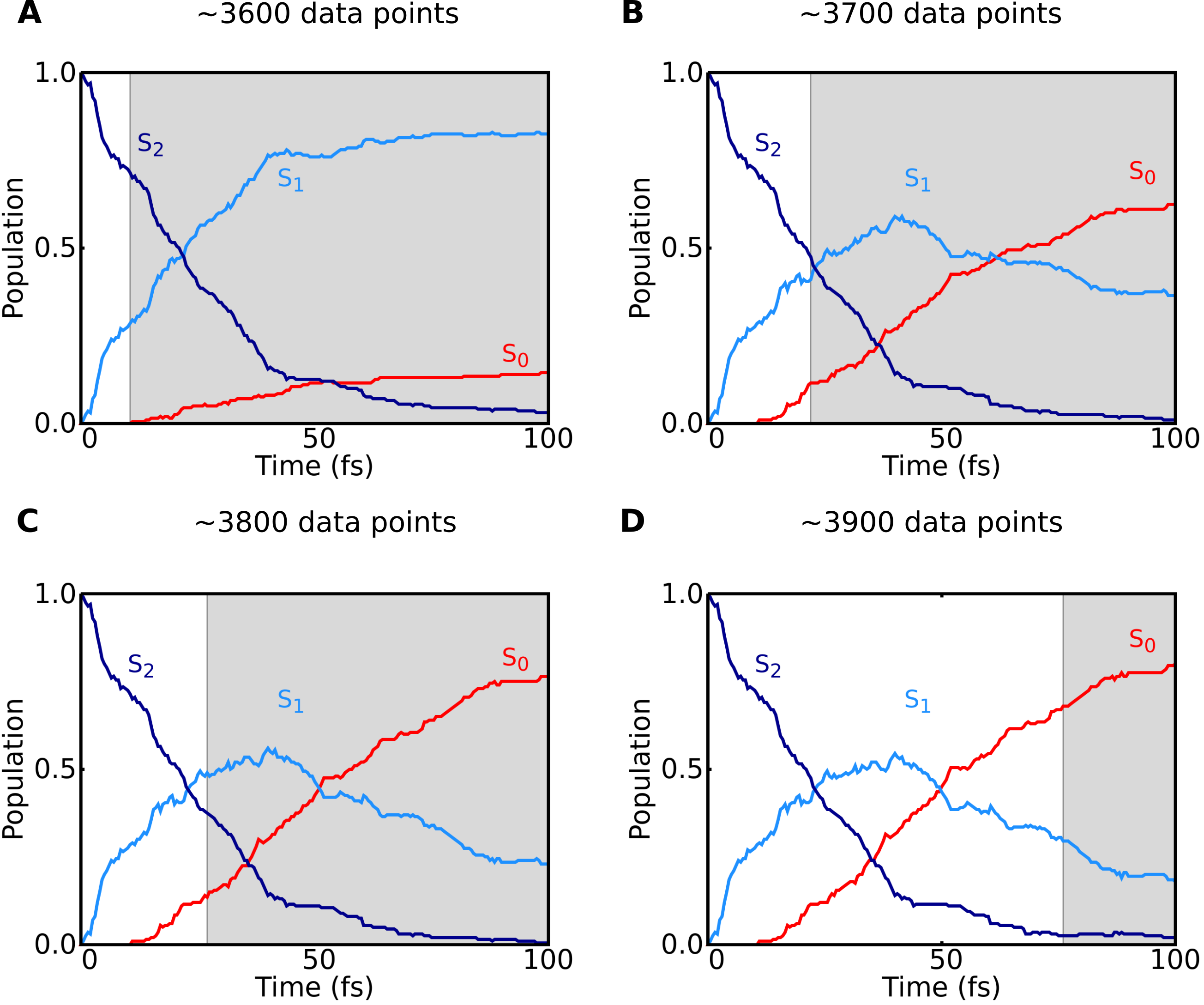}
    \caption{Predictions of dynamics with deep NNs during the last iterations of adaptive sampling. (A) Dynamics simulations with deep NNs trained on 3600 data points. The grey area indicates the time span, where trajectories were stopped by the adaptive sampling algorithm due to running into regions with insufficient training data. (B-D) Convergence towards the correct dynamics with NNs trained on approximately 3700, 3800, and 3900 data points. See Fig. \ref{fig:dynamics} for the reference dynamics. Stopping of trajectories by the sampling algorithm is shifted to later times with increasing training set size, as indicated by the grey areas.}
    \label{fig:adaptivesampling}
\end{figure}
 
\begin{table}[h]
\centering
\small
\caption{Selected parameters to construct NNs trained on data of CH$_2$NH$_2^+$ and carry out computations given in Figs. 3-6 of the main text and Figs. S4-6 of the ESI.~NN1 describes the first NN that was defined, whereas NN2 describes the second one. The influence of the forces for training of energies was set to 1.}
\label{tab:1}
\begin{tabular}{llll}
\hline
\hline
Property &Energy (NN1)  &Dipole moments (NN1) &NACs (NN1) \\
\hline 
Number of hidden layers &6&6&6\\
Number of neurons per hidden layer &50 &50 &74 \\
Batch size &500 &500 &500 \\
Learning rate, $lr$ &$5.33\cdot10^{-3}$ &$6.20\cdot 10^{-6}$&$5.54\cdot 10^{-5}$  \\
Decay factor, $f_{lr}$ &0.994	&0.9999	&0.994	\\
Update steps for annealing of $lr$ &3 &1 &64\\
L2 regularization rate &$3.32\cdot 10^{-8}$	&$8.63\cdot 10^{-7}$ &$3.99\cdot 10^{-8}$  \\
Basis function &Shifted softplus &Hyperbolic tangent &Hyperbolic tangent \\
\hline
Property &Energy (NN2)  &Dipole moments (NN2) &NACs (NN2)\\
\hline 
Number of hidden layers &6&6&6\\
Number of neurons per hidden layer &50 &50&79\\
Batch size &500 &500  &500\\
Learning rate, $lr$ &$3.49\cdot 10^{-3}$	&$2.74\cdot10^{-6}$ &$1.22\cdot 10^{-5}$\\
Decay factor, $f_{lr}$ &0.990	&0.993	&0.984\\
Update steps for annealing of $lr$ 	&5	&7	&79\\
L2 regularization rate &$5.33\cdot 10^{-7}$ &$2.39\cdot10^{-5}$	&$2.62\cdot10^{-8}$	\\
Basis function &Shifted softplus &Hyperbolic tangent &Hyperbolic tangent\\
\end{tabular}
\centering
\end{table}

\begin{table}[h]
\centering
\small
  \caption{Mean absolute error (MAE) and root mean-squared error (RMSE) of predicted properties averaged over all states for the training set of the methylenimmonium cation containing 4000 data points.}
 \label{tab:MAE_molecule}
\begin{tabular}{lllll}
     \hline
     \hline
     &Energy [H] &Gradients [H/Bohr] &Dipole moments [a.u.] &NACs [a.u.]  \\
     \hline
MAE &0.00117 &0.0189 &0.0971 &0.156\\
RMSE &0.00225 &0.0395 &0.208 &1.13\\
     \hline
\end{tabular}
\centering
\end{table}

As already mentioned, the predefined threshold is multiplied by a factor of 0.95 after a new conformation is detected, recalculated with QC and the NNs are retrained. The initial value was adapted by this factor after each training cycle for adaptive sampling, when the training set is still built.
For the long dynamics simulations (after 10 ps), two NNs are still used for detecting undersampled regions of conformational space but threshold for the RMSE between the NNs is now kept constant (at a value of 0.05 a.u. for energies). The training set size increases only slightly (by 58 samples) during this time and are not considered to lie within the previously visited regions of the PES. Those are considered to be sampled as dense as before and new data points are thus not supposed to change the performance of NNs in already passed regions of the PES.

\clearpage
\subsection{Performance of NNs}
In this subsection, we want to assess the performance of the generated NN potentials by comparing to different ML models and by an in-depth analysis of the predictions specific for each electronic state. We evaluate the accuracy by using ML models trained on the training set of the methylenimmonium cation computed with the \textit{QC1} method (see section \ref{sec:methylenimmonium} below). We want to point out here, that the QC reference dynamics simulations are not included in the training set and thus, reproducing the dynamics can be considered as an indirect test. However, we cannot use the points from the QC reference dynamics for a direct comparison in a straightforward manner, since these points are not phase corrected. In order to evaluate the accuracy of the NN potentials, we generated an out-of-sample test set, that contains energies and forces as well as NACs (and dipole moments for completeness) of 770 data points additionally generated via sampling of linear combinations of different normal modes of CH$_2$NH$_2^+$. This test set is phase corrected and is made available with the training set. We compare our NN model with another NN model using the Coulomb matrix~\cite{Rupp2012PRL} (referred to as NN/Coulomb in the following) with 21 features (inputs) instead of the inverse distance matrix (NN/inv.D. in the following) with 15 features as a molecular descriptor. We further generate polynomials of the inverse distance matrix and use them as descriptor (referred to as NN/poly.D.).  Therefore, we compute the outer product of two vectors containing the entries of the inverse distance matrix. The dimension of poly.D. is thus squared compared to inv.D., hence a more accurate description and higher accuracy could be expected. To compare between different ML models, we analyze the performance of two additional regression models: Linear regression (LR) computed as a baseline model and support vector machine for regression (SVR) is used for additional comparison. In both cases, the same molecular descriptor (inv.D.) as for the reference NNs is taken.

\subsubsection{Computational details for the ML models}
For training NN/Coulomb, LR and SVR models, we use the training set containing 4000 data points. We proceed in the same way as for the NN/inv.D. model described above and scale inputs and outputs for training and prediction according to equation \ref{eq:3}.
In the same manner as for the NN/inv.D. model, we sample different hyperparameters of the network architecture for the NN/Coulomb model and NN/poly.D. model (see discussion in section S1.1), but as a starting point we take optimal hyperparameters already found for the NN/inv.D. model (see Table \ref{tab:1}). Since no better performance is obtained using different hyperparameters for the NN/Coulomb model, we keep them unchanged and only change the molecular descriptor from the matrix of inverse distances to the Coulomb matrix~\cite{Rupp2012PRL}, $\mathbf{C}$, that additionally includes the atomic charges in the representation. The diagonal elements, $C_{ii}$, which are absent in the inv.D. descriptor, are constant values defined as
\begin{equation}
    \label{eq:coulomb}
     C_{ii} = \frac{1}{2}Z_i^{2.4}
\end{equation} 
and off-diagonal elements, $C_{ij}$, are computed as
\begin{equation}
    \label{eq:coulomb2}
    C_{ij} = \frac{Z_iZ_j}{\mid R_i -R_j\mid}.
\end{equation} 
$Z_i$ in equations \ref{eq:coulomb} and \ref{eq:coulomb2} refers to the nuclear charge of atom i and $R_i$ is the position of atom i. All electronic states are treated within one NN. Again, forces are trained together with potential energies and predicted as their derivatives. A separate NN is used for training of NACs.
In case of the NN/poly.D. model, a set of slightly different hyperparameters is chosen from hyperparameter search for energies and gradients and given in Table \ref{tab:param_poly}. The hyperparameters for nonadiabatic couplings are kept unchanged from NN/inv.D. models.

\begin{table}[h]
\centering
\small
\caption{Selected parameters to construct NN/poly.D. models trained on 4000 data points of CH$_2$NH$_2^+$. The parameter, $\eta$ in equation \ref{eq:2}, to control the influence of the forces for training of energies is set to 1. The parameters for NNs trained on NACs are depicted in Table \ref{tab:1}.}
\label{tab:param_poly}
\begin{tabular}{ll}
\hline
\hline
Property &Energy \\
\hline 
Number of hidden layers &6\\
Number of neurons per hidden layer &50 \\
Batch size &500  \\
Learning rate, $lr$ &$3.57\cdot10^{-3}$  \\
Decay factor, $f_{lr}$ &0.950	\\
Update steps for annealing of $lr$ &62 \\
L2 regularization rate &$6.10\cdot 10^{-8}$	 \\
Basis function &Shifted softplus \\
\end{tabular}
\centering
\end{table}

To carry out LR and SVR, scikit-learn~\cite{scikit-learn} is used. Each electronic state is treated separately and the forces and NACs are trained independently from the energies. Ordinary least squares LR is used as implemented in scikit-learn. For SVR, we tested different kernel types, mainly linear kernel, radial basis function and polynomials. For final training and prediction, radial basis functions are used. The penalty parameter of the error term is set to a value of 100 and the rest of the parameters are kept unchanged from default values, because the performance does not change considerably by setting different parameters. 
\\
\\
\subsubsection{Comparison of different ML models}
First, we seek to evaluate the performance of our NNs by comparison to other ML models. In our experiments, we carry out dynamics simulations with 2 NNs. Therefore, we compute the MAE averaged over all states for energies, forces, and NACs on the test set of 770 data points with two models. For the NN/inv.D. models, NN1 and NN2 are used, as they are in dynamics simulations. For the rest of the models, we split the data set of 4000 points in different training and validation sets using a ratio of 9:1 and use different initial weights. The results are given in Table \ref{tab:MAE_testset} (for completeness, the MAE on dipole moments for our model is 0.168 a.u., but will not be discussed here). As can be seen in the table, the NNs with different descriptors reach similar accuracy. Especially, the inverse distance matrix and the Coulomb matrix perform very similarly, as would be expected from such closely related descriptors. Due to the similarity, we keep the simplest of these descriptors (inv.D.) for further analysis. SVR is worse than the NN approaches with an error for energies and gradients approximately twice as large as for NNs. The SVR also delivers worse predictions of NACs compared to the NNs. LR, as a baseline model, has a MAE that is about a factor of 10 larger than the one of SVR for energies and gradients and about 1.5 times larger for NACs, demonstrating the utility of our NN approach described above. 

\begin{table}[h]
\centering
\small
  \caption{The mean absolute error (MAE) averaged over all states for energies, gradients, and NACs on a test set of 770 data points of the methylenimmonium cation for linear regression models, LR, support vector machines for regression, SVR, and NNs with the inverse distance (inv.D.), Coulomb matrix (Coulomb), and inverse distance polynomials (poly.D.) as molecular descriptors.}
 \label{tab:MAE_testset}
\begin{tabular}{llll}
     \hline
     \hline
     Model & MAE Energy [H] & MAE Gradients [H/Bohr] & MAE NACs [a.u.]  \\
     \hline
     NN/inv.D.   & 0.00237 & 0.00669 & 0.328  \\
     NN/Coulomb  & 0.00238 & 0.00690 & 0.314 \\
     NN/poly.D.  & 0.00197 & 0.00617 & 0.335        \\
     LR/inv.D.   & 0.09240  & 0.13902   & 0.471 \\
     SVR/inv.D.  & 0.00618 & 0.01169  & 0.382 \\ 
     \hline
     \hline
     \end{tabular}
\centering
\end{table}

\subsubsection{State-specific analysis of the NNs}
In order to clarify whether there is a bias of our ML model toward a specific state, we evaluate the errors of our reference NN/inv.D. model on energies and gradients on the same test set of 770 data points for each electronic state separately. As can be seen in Table ~\ref{tab:MAE_testset_states}, the MAE of the S$_2$-state energies and gradients is approximately twice as large as the one of the two lower states. Analysis of different reaction coordinates of the PES reveal problems within QC calculations that become pronounced especially in critical regions of the PES and lead to erratic potential energy curves. To show this problem, a scan along a reaction coordinate that includes two avoided potential energy curve crossings of the molecule is exemplified in Fig.~\ref{fig:scan}. As can be seen, a jump in the S$_2$ potential energy curve is obtained with the \textit{QC1} reference method. The source of this behaviour might be due to higher electronic states that enter along the reaction coordinate ("intruder states") or new electronic configurations of the molecule that were not included in the active space from the beginning. However, the NNs used in this work (panel A in Fig. \ref{fig:scan}) do not reproduce this discontinuity and predict smooth potential energy curves, certainly due to the use of gradients in the loss function for the training. As a consequence, the MAE obtained for the S$_2$ state must be artificially higher than for the rest of the states.

\begin{table}[h]
\centering
\small
  \caption{The MAE on energies and gradients of each electronic state on the test set of 770 data points obtained with the NNs used in this work to carry out dynamics simulations.}
 \label{tab:MAE_testset_states}
\begin{tabular}{ccc}
     \hline
     \hline
     State & MAE Energy [H] & MAE Gradients [H/Bohr] \\
     \hline
      S$_0$  & 0.00176 & 0.00444 \\
      S$_1$  & 0.00200 & 0.00670 \\
      S$_2$  & 0.00335 & 0.00893 \\
     \hline
     \hline
\end{tabular}
\centering
\end{table}

For completeness, we carry out the same scan with the other ML models and as expected,  the potential energy curves obtained with NN/Coulomb (panel B in Fig. \ref{fig:scan}) are very similar to the ones obtained with NN/inv.D. (panel A) and NN/poly.D. (panel C). The same trend as in Table \ref{tab:MAE_testset} is also obtained for the other models. The SVR model (panel E) gets the potential energy curves slightly worse than the NNs and LR is far off from the reference potential energy curves (panel D).

\begin{figure}[h]
    \centering
    \includegraphics[height=9cm]{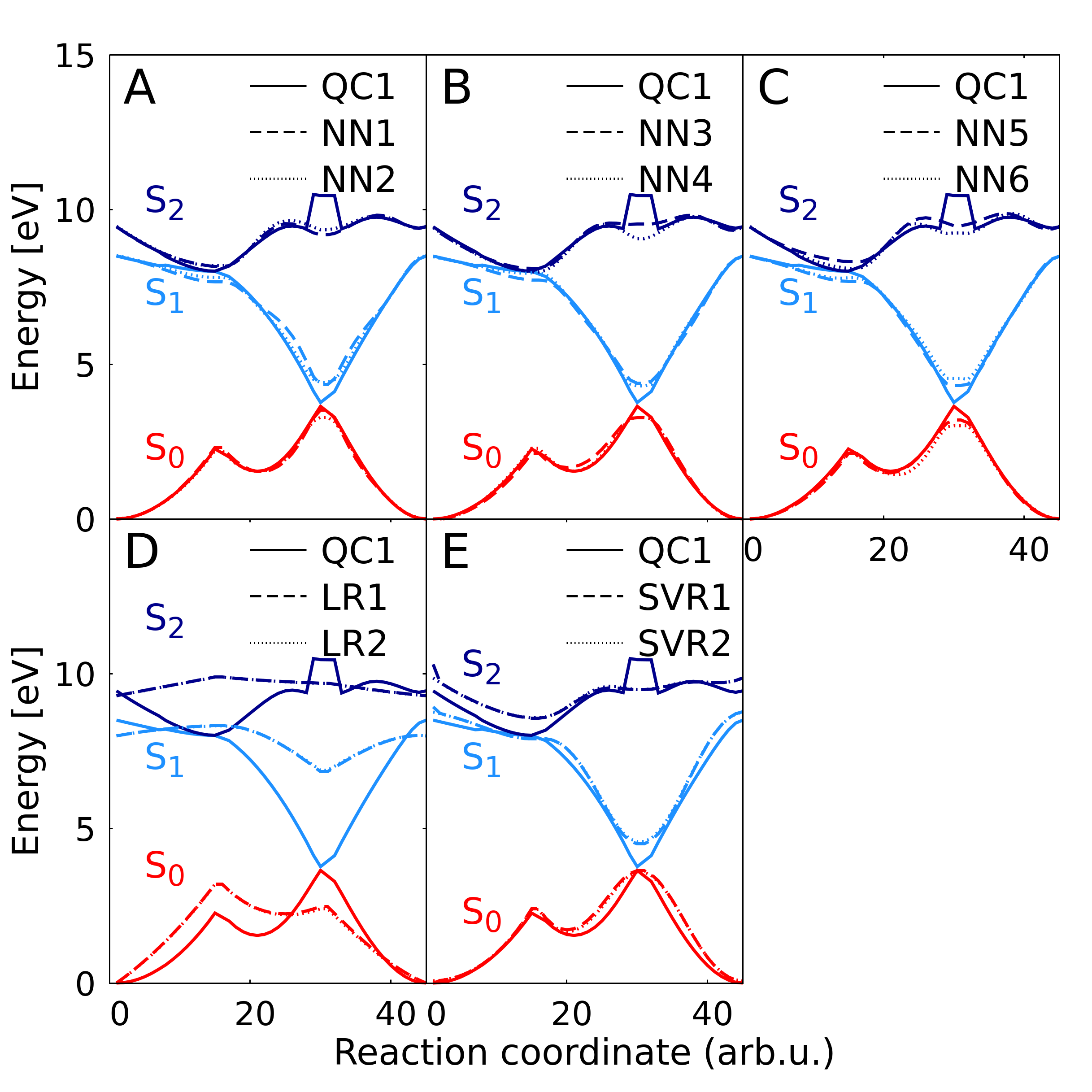}
    \caption{Scan along a reaction coordinate of CH$_2$NH$_2^+$ showing two avoided crossings, one between the S$_2$ and S$_1$ state and one between the S$_1$ and S$_0$ state computed with QC1 (continuous line) and different ML models (dashed and dotted lines). (\textbf{A}) NN1 and NN2 are the reference NNs used in this work (NN/inv.D.), ( \textbf{B}) NN3 to NN6 are again multi-layer feed-forward NNs, but with the Coulomb matrix~\cite{Rupp2012PRL} (NN3 and NN4: NN/Coulomb), (\textbf{C}) and polynomials of the inverse distances (NN5 and NN6: NN/poly.D.) as molecular descriptors.  (\textbf{D}) LR1 and LR2 indicate linear regression models, (\textbf{E}) and SVR1 and SVR2 are support vector machines for regression. In case of the \textit{QC1} calculation, erratic behaviour of the S$_2$ potential energy curve can be seen.}
    \label{fig:scan}
\end{figure}

An important feature of the ML models used here is their tendency towards a smooth interpolation of the potential energy surfaces (see Fig. S2). As a consequence, high artificial errors and discontinuities introduced by the reference method can be compensated to a certain extent. Moreover, since all ensemble models predict the same trend in such situations (see e.g. NN1 and NN2 Fig. S2A), a low uncertainty is reported for the afflicted regions and no potentially detrimental data is added to the data set during the adaptive sampling. This feature is desired and advantageous since
the ML models correct for the erratic behavior of QC1.

Nevertheless, such regions often coincide with critical regions of the PES and should be considered with care. To highlight such regions in conformational space that should be sampled more comprehensively in the training set, some other metrics could be considered. For example, very small energy gaps between different states or large coupling values are an indicator for a crossing point of two potentials. 
Additionally, the root-mean squared displacement (RMSD) from a new configuration to all other configurations contained in the training set could be calculated and a threshold could be set, which tells us when a new geometry is dissimilar to the current training set and should be computed with QC and added to the training set. Another possibility would be to add more data points from trajectories where the gradient is large, which is done in Ref.~\cite{Pukrittayakamee2009JCP} for example for the generation of a database. This is important, because during dynamics simulations the molecule will remain longer in regions where the forces are small, hence enlarging the training set mainly by data points with small gradients.~\cite{Pukrittayakamee2009JCP} Another procedure is used for Shepard interpolation.~\cite{Thompson1997JCSFT} There, a statistical approach with no significant additional costs is used to determine the uncertainty of the potential energy by measuring its variance.

\clearpage
\section{Phase correction}
We carried out a phase correction procedure to obtain a training set suitable for our ML algorithm. To this aim, a phase vector, $\mathbf{p}$, had to be derived for each molecular geometry containing values of +1 and -1 for each state. This vector was obtained by the computation of wave function overlaps between two different geometries. The overlap matrix, $\mathbf{S}$, between wave functions of following nuclear configurations, $\Psi_k$ and $\Psi_l$, was computed using the WFoverlap code:\cite{Plasser2016JCTC}
\begin{equation}
\label{eq:wf}
    \mathbf{S}=\langle\Psi_k\mid\Psi_l\rangle.
\end{equation}
The overlap matrix has the dimension of $N^{states} \times N^{states}$, with $N^{states}$ being the number of electronic states included in a calculation. When the molecular geometries are similar enough to each other, overlaps are sufficiently large to gain information whether a phase switch has occurred or not. Wave function overlaps for an electronic state that are close to +1 indicate that no phase change has taken place, whereas overlaps close to -1 point out a change in the phase of a wave function. Far from conical intersections, $\mathbf{p}$ usually corresponds to the diagonal matrix elements of $\mathbf{S}$ and also has the dimension of $N^{states}$. Whenever two PESs are in close proximity, off-diagonal elements of $\mathbf{S}$ can be larger than diagonal elements. In such cases, values of $\mathbf{p}$ depend on absolute values of row elements of $\mathbf{S}$. A threshold of $\pm0.5$ was set between overlaps of two following wave functions to ensure that the correct value of +1 or -1 was taken into account. Each entry, $p_i$, of $\mathbf{p}$ was then calculated according to Equation \ref{eq:phase1} with i and j running over all $N^{states}$. 
\begin{equation}
    \label{eq:phase1}
    p_i = sgn(max(\mid S_{ij}\mid)sgn(S_{ij})) \forall \mid S_{ij}\mid\geq0.5; i,j=1,2,...,N^{states}
\end{equation}

However, many situations arise, in which computed overlaps between two geometries yield values close to 0. With these, it is not possible to decide reliably whether a phase change is present or not. Therefore, an interpolation of n steps between those two molecular geometries needed to be carried out, with n usually being in the range of 5 to 10. Interpolation was performed with respect to a molecule's Z matrix. Transformation from xyz format to a Z matrix format and vice versa was executed with OpenBabel.\cite{OpenBabel2011} The size of interpolation steps was sufficient when each computation yielded overlaps close to +1 or -1. The phase vector, $\mathbf{p}_n$, applicable to the last point, $n$, of the required interpolation was found by multiplication of all previous phase vectors, $\mathbf{p}_0$ to $\mathbf{p}_{n-1}$:

\begin{equation}
    \label{eq:phase2}
    \mathbf{p_n}=\prod_{\beta=0}^{n-1}\mathbf{p}_\beta. 
\end{equation}
$\mathbf{p}_0$ always corresponds to the phase vector between the reference geometry, upon which the complete training set is phase corrected, and the first interpolation step. The reference geometry for global phase correction was chosen to be the equilibrium geometry.  

If no row element of $\mathbf{S}$ had an absolute value above or equal to $\pm$0.5, the interpolation was stopped and more interpolated nuclear configurations between the equilibrium geometry and the one that needs to be included in the training set were generated. If there were still not enough interpolated configurations to account for sufficiently large overlaps between two subsequent wave functions, the simulations were stopped. One reason for too small overlaps are "intruder states", which are excluded at the reference geometry, but are included at another geometry due to an energy drop. Thus, a previously included state is excluded such that the overall molecular electronic wave function changes considerably.

The phase correction of each matrix $\mathbf{M}$ -- in our case, the Hamiltonian matrix, which contains energies and couplings, and the dipole matrices for each direction of the coordinate system -- was carried out according to Equation \ref{eq:phase3}. Matrices containing all relevant nonadiabatic coupling vectors to form the nonadiabatic coupling tensor with the dimension $N^{states}xN^{states}xN^{atoms}x3$ (with $N^{atoms}$ being the number of atoms) were corrected according to Equation \ref{eq:phase4}. Each two dimensional nonadiabatic coupling vector, $\mathbf{NAC}^{ij}$, accounts for NACs between two states i and j. It was corrected by multiplication of every element with the entry of $\mathbf{p}$ for each state i, ${p}_i$, and j, ${p}_j$.

\begin{equation}
    \label{eq:phase3}
    \mathbf{M} = M_{ij}\cdot p_i\cdot p_j; i,j = 1,2,...,N^{states}
\end{equation}

\begin{equation}
    \label{eq:phase4}
    \mathbf{NAC}^{ij}=\mathbf{NAC}^{ij}\cdot p_i \cdot p_j
\end{equation}

\clearpage
\section{Analytical Model}

In order to check the performance of our deep learning molecular dynamics approach we constructed an one-dimensional, diabatic analytical model consisting of five harmonic oscillators, with the analytical interface of SHARC.\cite{Mai2018WCMS,sharc-md2} The potential energy curves are defined in Table \ref{tab:Analytical_PEC}. Nonadiabatic potential couplings (in contrast to nonadiabatic derivative couplings) are responsible for transitions between the $^1X$ and $^1A$ state while spin-orbit couplings coupled states of different spin multiplicity, i.e. $^1X$ and $^3B$ as well as $^1A$ and $^3B$. These couplings as well as dipole moments between different states are described by constant values given in Table \ref{tab:Analytical_Coupling} and \ref{tab:Analytical_Dipole}, respectively. A total number of 2000 initial starting points were sampled according to a Wigner distribution,\cite{Mai2018WCMS} from which 100 were selected, excited in a range of 0-2 eV and propagated for 100 fs with a time step of 0.05 fs for the nuclear motion and 0.001 fs for the analytical solution of electronic amplitudes. 

The five harmonic oscillators of the analytical model are given in Fig. \ref{fig:Analytical} and represent the ground state $^1X$ (dark blue line in Fig. \ref{fig:Analytical}A), one excited singlet state $^1A$ (light blue line in Fig. \ref{fig:Analytical}A), and three degenerate excited states to mimic a triplet state, $^3B$ (red line in Fig. \ref{fig:Analytical}A). The states are coupled by both nonadiabatic potential couplings and spin-orbit couplings. Sampling of configurations along the degree of freedom leads to a training set of 100 data points and the potential energy curves can be fitted nearly exactly by NNs (solid lines in Fig. \ref{fig:Analytical}A). Excited-state dynamics are started from 100 different initial conditions generated in the $^1X$ state, each excited to the $^1A$ state. The time evolution of the different states using the original analytical potentials and the trained NNs (Fig. \ref{fig:Analytical}B) show that the predictions are comparable and each state is populated similarly after 100 fs. The MAE in the predictions of the energies is 0.00031 eV (0.01 kcal/mol) and of the gradients is 0.0024 eV/\AA ~(0.057 kcal/mol/\AA). Hyper parameters  chosen for the NNs can be obtained from Table \ref{tab:Analytical_parameter}. This first proof-of-concept thus shows the ability of our ML method to describe excited-state dynamics including nonadiabatic potential and spin-orbit couplings.

\begin{figure}[h]
    \centering
    \includegraphics[height=5cm]{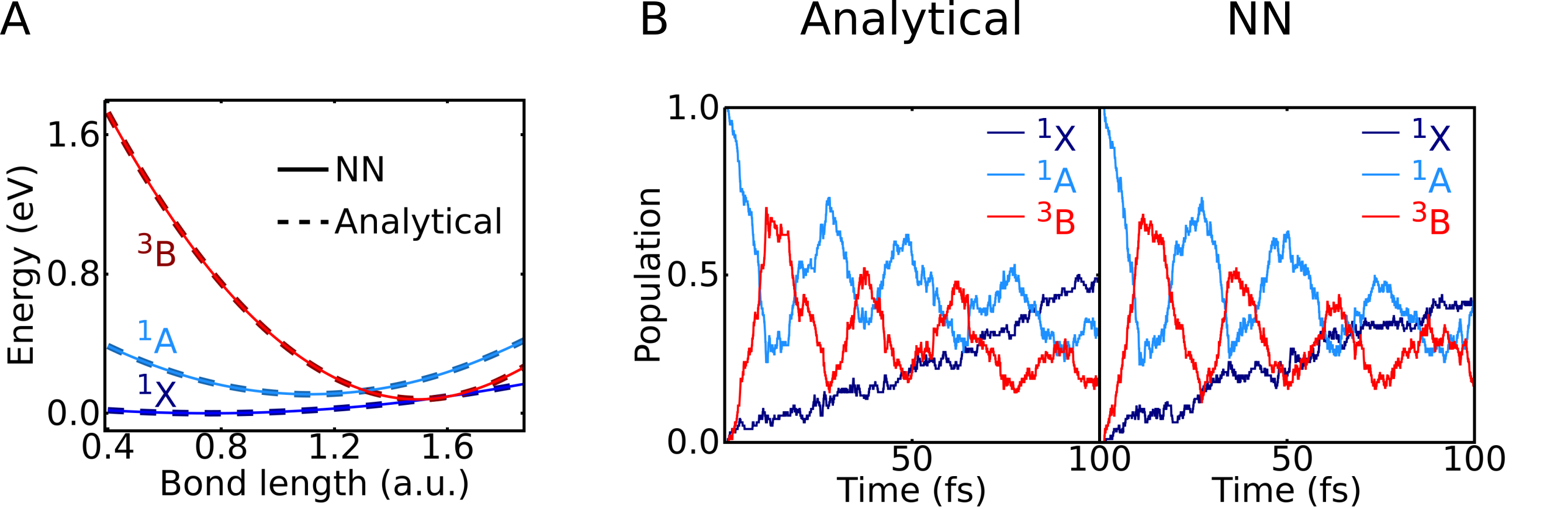}
    \caption{\textbf{Analytical model}. \textbf{(A)} Analytical (dashed lines) and neural network (NN, solid lines) potential energy curves along an one-dimensional model. \textbf{(B)} Time evolution of the different electronic state population after excitation from the $^1X$ state to the $^1A$ state using the exact analytical setup and NNs.}
    \label{fig:Analytical}
\end{figure}

\begin{table}[h]
\centering
\small
\caption{Parameters of harmonic potentials according to the equation $V(x)=0.5k(x-x_0 )^2+E_0$.}
\label{tab:Analytical_PEC}
\begin{tabular}{llll}
\hline
\hline
Energetic state &k[a.u.] &$x_0$ [a.u.] &$E_0$ [a.u.]\\
\hline
$^1X$ &0.01 &0.756 &0.0 \\
$^1A$ &0.04 &1.115 &0.004 \\
$^3B$ &0.1 &1.50 &0.003\\
\hline 
\end{tabular}
\centering
\end{table}

\begin{table}[h]
\centering
\small
\caption{Couplings between different electronic states of the analytical model.}
\label{tab:Analytical_Coupling}
\begin{tabular}{llllll}
\hline
\hline
Couplings between states &$^1X$ [a.u.] &$^1A$ [a.u.] &$^3B(+)$ [a.u.] &$^3B(0)$ [a.u.] &$^3B(-)$ [a.u.]\\
\hline
$^1X$ && 0.00025 &0.005-0.0001i &-0.0025i &0.005+0.0001i \\
$^1A$ &0.00025 & &0.001+0.0025i &+0.0025i &0.001-0.0025i\\
$^3B(+)$ &0.005+0.0001i &0.001-0.0025i & & & \\
$^3B(0)$ &0.0025i &-0.0025 i&&&\\
$^3B(-)$ &0.005-0.0001i &0.001+0.0025i\\
\hline 
\end{tabular}
\centering
\end{table}

\begin{table}[h]
\centering
\small
\caption{Dipole moments between different electronic states of the analytical model.}
\label{tab:Analytical_Dipole}
\begin{tabular}{llllll}
\hline
\hline
Dipole moment between states &$^1X$ [a.u.] &$^1A$ [a.u.] &$^3B(+)$ [a.u.] &$^3B(0)$ [a.u.] &$^3B(-)$ [a.u.]\\
\hline
$^1X$ &-1.90 &-0.3 \\
$^1A$ &-0.3 &-1.2 \\
$^3B(+)$ & & &-0.3 & &\\
$^3B(0)$ & & & &-0.3 &\\
$^3B(-)$ & & & & &-0.3\\
\hline 
\end{tabular}
\centering
\end{table}

\begin{table}[h]
\centering
\small
\caption{Selected parameters to construct NNs trained on data of the analytical model to carry out dynamics simulations given in Fig. \ref{fig:Analytical}. NN1 describes the first NN that was defined, whereas NN2 describes the second one. The influence of the forces for training of energies was set to 1.}
\label{tab:Analytical_parameter}
\begin{tabular}{llll}
\hline
Property &Energy (NN1) &Couplings (NN1) &Dipole moments (NN1) \\
\hline 
Number of hidden layers &1 &2 &2\\
Number of neurons per hidden layer &13 &32 &36\\
Batch size &50 &50 &50 \\
Learning rate, $lr$ &$4.83\cdot10^{-3}$ &$9.37\cdot 10^{-4}$ &$9.41\cdot 10^{-4}$ \\
Decay factor, $f_{lr}$ &- &-&-	\\
Update steps for annealing of $lr$ &- &- &-\\
L2 regularization rate &$6.07\cdot 10^{-10}$	&$2.35\cdot 10^{-12}$ &$6.23\cdot 10^{-10}$ \\
Basis function &Shifted softplus function &Hyperbolic tangent &Hyperbolic tangent \\
\hline 
\hline
Property &Energy (NN2) &Couplings (NN2) &Dipole moments (NN2)\\
\hline 
Number of hidden layers &1&2&2\\
Number of neurons per hidden layer &39 &32&24\\
Batch size &50 &50  &50\\
Learning rate, $lr$ &$7.89\cdot 10^{-3}$ &$9.31\cdot 10^{-4}$	&$9.48\cdot10^{-4}$\\
Decay factor, $f_{lr}$ &- &- &-\\
Update steps for annealing of $lr$ 	&- &- &-\\
L2 regularization rate &$6.23\cdot 10^{-10}$	&$1.94\cdot10^{-12}$	&$1.50\cdot10^{-14}$\\
Basis function &Shifted softplus function &Hyperbolic tangent &Hyperbolic tangent t\\
\end{tabular}
\centering
\end{table}

\clearpage
\section{Methylenimmonium cation}\label{sec:methylenimmonium}

\subsection{Computational details for QC}
The quantum chemical reference calculations of the methylenimmonium cation, CH$_2$NH$_2^+$, were done with the program COLUMBUS\cite{Lischka2001PCCP} which uses the implementation of nonadiabatic coupling vectors via Dalton\cite{Helgaker1997} integrals. CH$_2$NH$_2^+$ possesses 16 electrons. Molecular orbitals were optimized with the state-averaged complete active space self-consistent field (SA-CASSCF) method, averaged over 3 states. The CAS was set to 6 active electrons in 4 active orbitals. The two lowest molecular orbitals representing the 1s orbitals of the C and N atom (Fig. \ref{fig:orbitals}A) were frozen at the MRCI level of theory. The next three lowest energetic orbitals were set to be inactive (Fig. \ref{fig:orbitals}B) and 4 orbitals were assigned to the reference space with 6 active electrons (Fig. \ref{fig:orbitals}C). Single and double excitations from the (6,4) reference space to 73 virtual orbitals yield a total of about 660000 configuration state functions. Determinants from inactive orbitals to virtual orbitals for single and double excitation are included within MRCI-SD. The frequency calculation was carried out with COLUMBUS.
We performed scans with 100 points along each normal mode (Table \ref{tab:normalmode}) as well as with 72 points along the torsion around the central double bond in order to generate the initial training set. From these scans, we removed data points from computations that did not show proper convergence. After adaptive sampling, we ended up with a training set of 4000 data points. 

\begin{figure}[h]
    \centering
    \includegraphics[height=9cm]{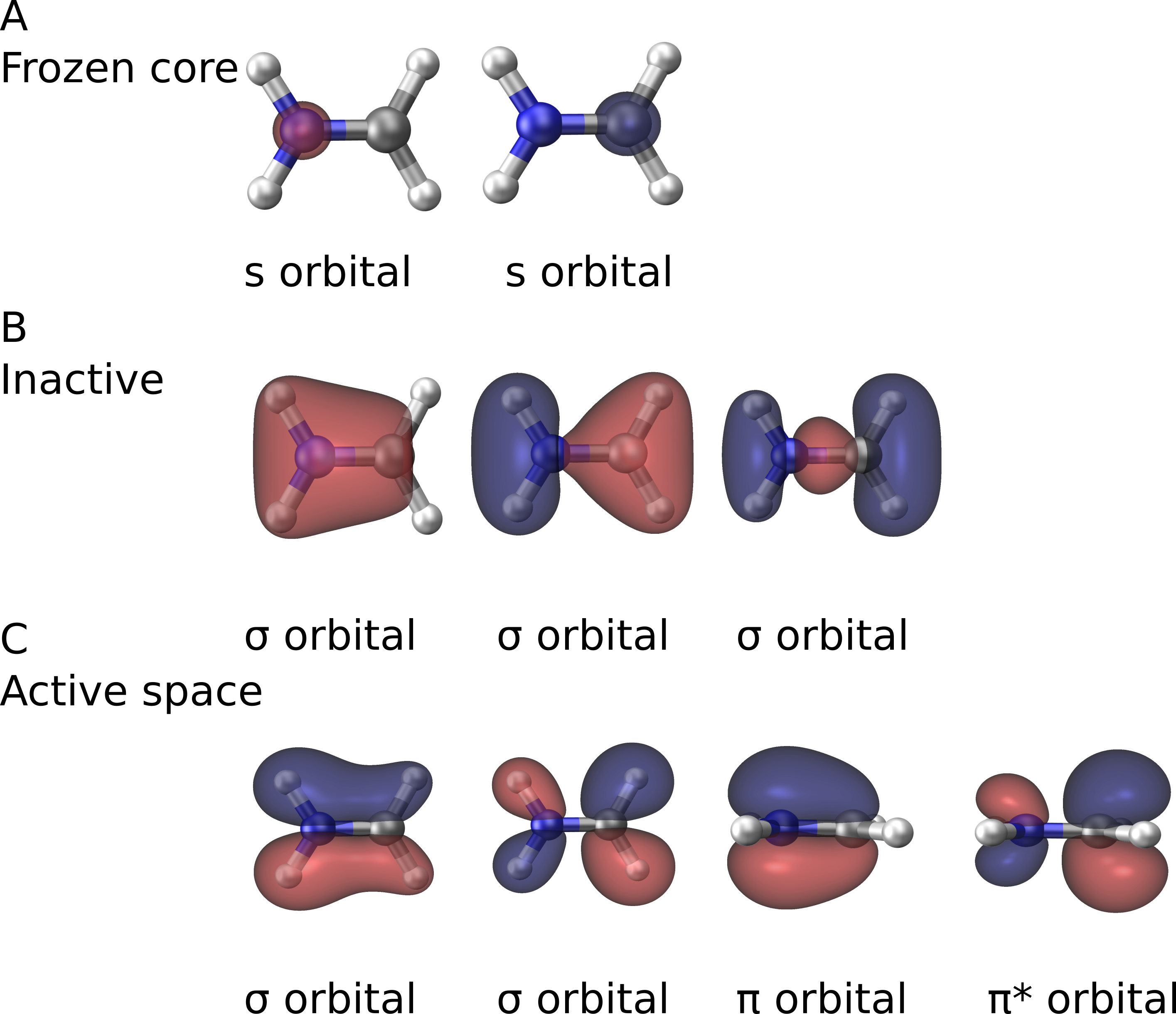}
    \caption{\textbf{Isosurface plots of the orbitals of the methylenimmonium cation.} \textbf{(A)} Orbitals representing the frozen core. \textbf{(B)} Orbitals representing inactive doubly occupied space energetically between the frozen core and the active space. \textbf{(C)} The active space consisting of 6 electrons in 4 orbitals.}
    \label{fig:orbitals}
\end{figure}

\begin{table}[h]
\centering
\small
\caption{Frequencies of each normal mode of CH$_2$NH$_2^+$ computed with MR-CISD/aug-cc-pVDZ.}
\label{tab:normalmode}
\begin{tabular}{ll}
\hline
\hline 
Degree of freedom (normal mode) &Frequency [$cm^{-1}$]\\
\hline 
1 & 940.09\\
2 & 969.35\\
3 & 1074.45\\
4 & 1174.32\\
5 & 1368.83\\
6 & 1473.26\\
7 & 1612.35\\
8 & 1789.73\\
9 & 3223.36\\
10 &3357.28 \\
11 &3550.29\\
12 &3663.38\\
\end{tabular}
\centering
\end{table}

\subsection{Surface-hopping molecular dynamics}

As a prerequisite for the dynamics, 1000 initial conditions were sampled from the Wigner distribution of the quantum harmonic oscillator defined by the above-mentioned frequency calculation. From these 1000 possible starting points, 200 were excited -- according to the oscillator strengths in the excitation window of 9.44 $\pm$ 0.15 eV -- to the brightest state, which is the second excited singlet state. This excitation shows mostly $\pi\pi^{\ast}$ character. Each trajectory was propagated for 100 fs with a time step of 0.5 fs for nuclear motion and 0.02 fs for integration of the time-dependent Schr\"{o}dinger equation. Trajectories showing problems with energy conservation due to improper convergence of the quantum chemistry were excluded, resulting in 90 trajectories for \textit{QC1} and 88 trajectories for \textit{QC2}. However, the trend of the populations when all trajectories were taken into account is the same. Therefore, no bias was introduced by sorting out trajectories due to bad energy convergence within quantum chemical calculations, which appeared mainly around conical intersections due to convergence problems. Such problems do not occur in the NNs, which demonstrates another advantage of the ML approach. Fig. \ref{fig:dynamics}A depicts a direct comparison of the molecular dynamics computed with \textit{QC1} (MR-CISD/aug-cc-pVDZ) and NNs. In both cases, the populations of 90 trajectories starting from the second excited singlet state are shown. Two hundred initial conditions were excited with 9.44 $\pm$ 0.15 eV and 90 trajectories reached 100 fs in the case of \textit{QC1}. Therefore, from NN-simulations only the first 90 trajectories were used for comparison. As can be seen, the populations of each state are in good agreement when comparing both methods -- as they are in Fig. 3 in the main text. In the case of the populations shown in Fig. 4 in the main text, we propagated 200 trajectories up to 10 ps. Afterwards, for demonstrating the possibility of long time scale simulations, we propagated 2 trajectories up to 1 ns. The populations up to 10 ps are averaged over 200 trajectories, the populations from 10 ps on up to 1 ns are averaged over 2 trajectories, respectively. In case of the nanosecond time scale simulations, every 100$^{th}$ time step, equivalent to every 50 fs, is written out by the program pySHARC. The kinetic constants discussed in the main text are obtained using the tools of SHARC.~\cite{sharc-md,Mai2018WCMS} By solving differential equations that describe the kinetics of the underlying model, a fit for the population transfer can be obtained and rate constants can be derived.

In addition to the population dynamics, we also checked whether visited molecular geometries are comparable along the trajectories. We first calculated the mean of each nuclear configuration over time from the NNs leading to populations given in Fig. \ref{fig:dynamics} and computed the RMSD to the mean molecular geometry at the respective time step of the ensemble predicted with MR-CISD/aug-cc-pVDZ, i.e. \textit{QC1} (Fig. \ref{fig:dynamics}B continuous line). An analogous comparison is also carried out for \textit{QC1} and \textit{QC2} (MR-CISD/6-31++G**, Fig. \ref{fig:dynamics}B, dotted line). For completeness, we also computed the RMSD between NN and \textit{QC2} (Fig. \ref{fig:dynamics}B, dashed line). All RMSDs are of comparable size, further validating our NN approach. 

\begin{figure}[h]
    \centering
    \includegraphics[height=6cm]{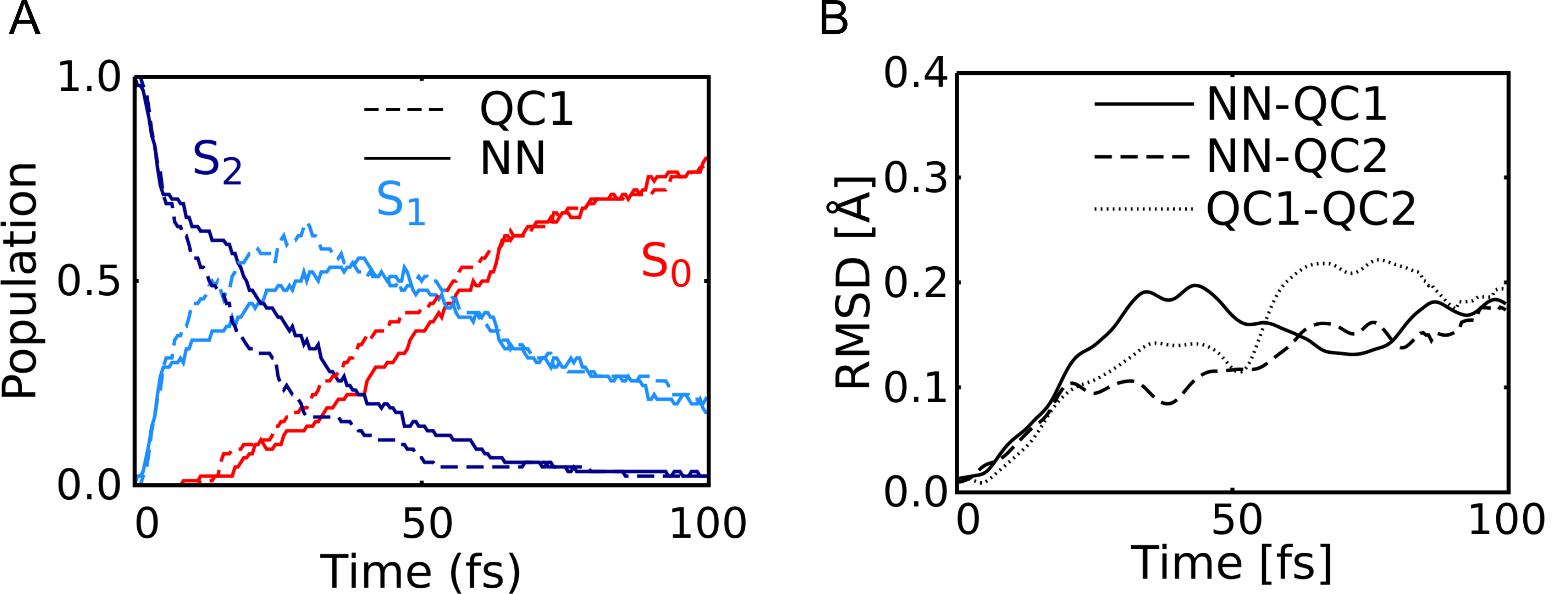}
    \caption{\textbf{(A)} Comparison of dynamics from \textit{QC1} and NN with 90 trajectories, respectively. The continuous lines show the populations of each excited singlet state calculated with a deep neural network (NN), whereas the  dashed line shows the corresponding population computed with \textit{QC1} (MR-CISD/aug-cc-pVDZ). \textbf{(B)} RMSD of the mean of nuclear configurations of each trajectory over time. NN-\textit{QC1} is the RMSD between deep NNs and the reference quantum chemical method \textit{QC1} (MR-CISD/aug-cc-pVDZ), NN-\textit{QC2} is the RMSD between deep NNs and the \textit{QC2} method (MR-CISD/6-31++G**), whereas \textit{QC1}-\textit{QC2} is the RMSD between \textit{QC1} and \textit{QC2}.}
    \label{fig:dynamics}
\end{figure}

Further, we checked the energy conservation of trajectories with the reference method \textit{QC1} and NNs independently. For both methods, we computed the mean and standard deviation (Std.) of the total energy along a trajectory, where we once allow hopping and once force the trajectories to stay in the S$_2$ state after excitation. Results are shown in Table \ref{tab:energyconservation}. As can be obtained, the mean total energy as well as the Std. along trajectories propagated for 100 fs are comparable among the methods. However, the Std. increases slightly when hops are forbidden between different energetic states. This trend is obtained with both methods. Moreover, when a time step of 0.05 fs is used instead of 0.5 fs for classical propagation of the nuclei, the energy conservation improves, which meets expectations. Worth mentioning is that, in the case of \textit{QC1}, trajectories which show large steps in total energy or do not reach 100 fs at all are removed prior to analysis. This is not the case for NNs, as all trajectories were suitable for analysis and no prior selection was carried out.

\begin{table}[h]
\centering
\small
  \caption{MAE and standard deviation (Std.) of the total energy obtained with the \textit{QC1} (MR-CISD/aug-cc-pVDZ) method and neural networks (NN). In the case of a time step of 0.5 fs for classical propagation of the nuclei, 90 trajectories were analyzed, whereas 50 trajectories were used for analysis of trajectories with a time step of 0.05 fs. }
 \label{tab:energyconservation}
\begin{tabular}{lllll}
\hline
\hline
Method &Time step [fs] &MAE [eV] &Std. [eV] &Hops allowed?\\
\hline
\textit{QC1} &0.5 &10.63 &0.047 &Yes\\
\textit{QC1} &0.5 &10.77 &0.059 &No\\
\textit{QC1} &0.05 &10.73 &0.011 &No\\
NN &0.5 &10.72 &0.052 &Yes\\
NN &0.5 &10.73 &0.061 &No\\
NN &0.05 &10.80 &0.017 &No\\
\hline
\end{tabular}
\centering
\end{table}

\subsection{Conical intersections (CIs)}
With each method, \textit{QC1}, \textit{QC2}, and NNs, we optimized the geometries of the two S$_1$/S$_0$ and S$_2$/S$_1$ CIs. Optimizations were executed with the SHARC tools that utilize an external optimizer of ORCA,\cite{Neese2012WCMS} where the computed energies and gradients\cite{Levine2007JPCB,Bearpark1994CPL} from COLUMBUS or NNs are fed in. Geometries are shown in Fig. \ref{fig:CIgeom}. They were optimized starting from the different hopping geometries obtained with the afore-mentioned methods, as indicated in the scatter plot of Fig. 6 in the main text and the zoom in Fig. \ref{fig:CI}. The optimized molecular geometries agree well. Cartesian coordinates are given in Table \ref{tab:CifirstNN}-\ref{tab:CilastQC}.

\begin{figure}[h]
    \centering
    \includegraphics[height=6cm]{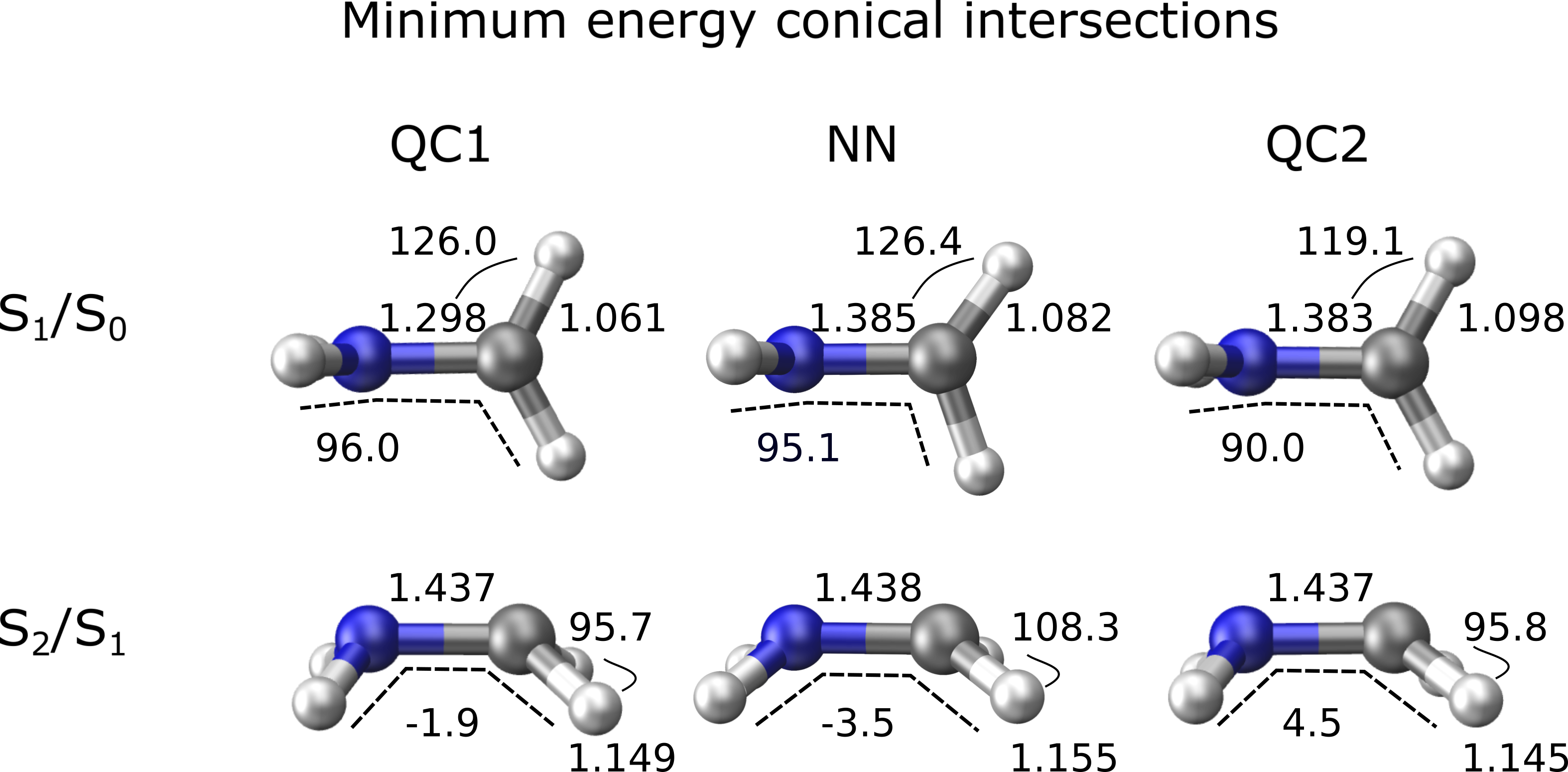}
    \caption{\textbf{Conical intersections (CIs).} For each method, we could find two different CIs. \textit{QC1} (MR-CISD/aug-cc-pVDZ) is used as the reference method to train neural networks (NNs) and compared to CIs obtained with \textit{QC2} (MR-CISD/6-31++G**). For the two CIs, the bond length between the nitrogen (blue) and the carbon (grey) atom is shown, as well as the bond length between the carbon and one hydrogen atom. Values are given in angstrom. A dihedral angle between four atoms marked with the dashed line is given, as well as an angle between the carbon and a hydrogen atom (S$_1$/S$_0$ CI) and between two hydrogen atoms (S$_2$/S$_1$ CI).}
    \label{fig:CIgeom}
\end{figure}

\begin{figure}[h]
    \centering
    \includegraphics[height=7cm]{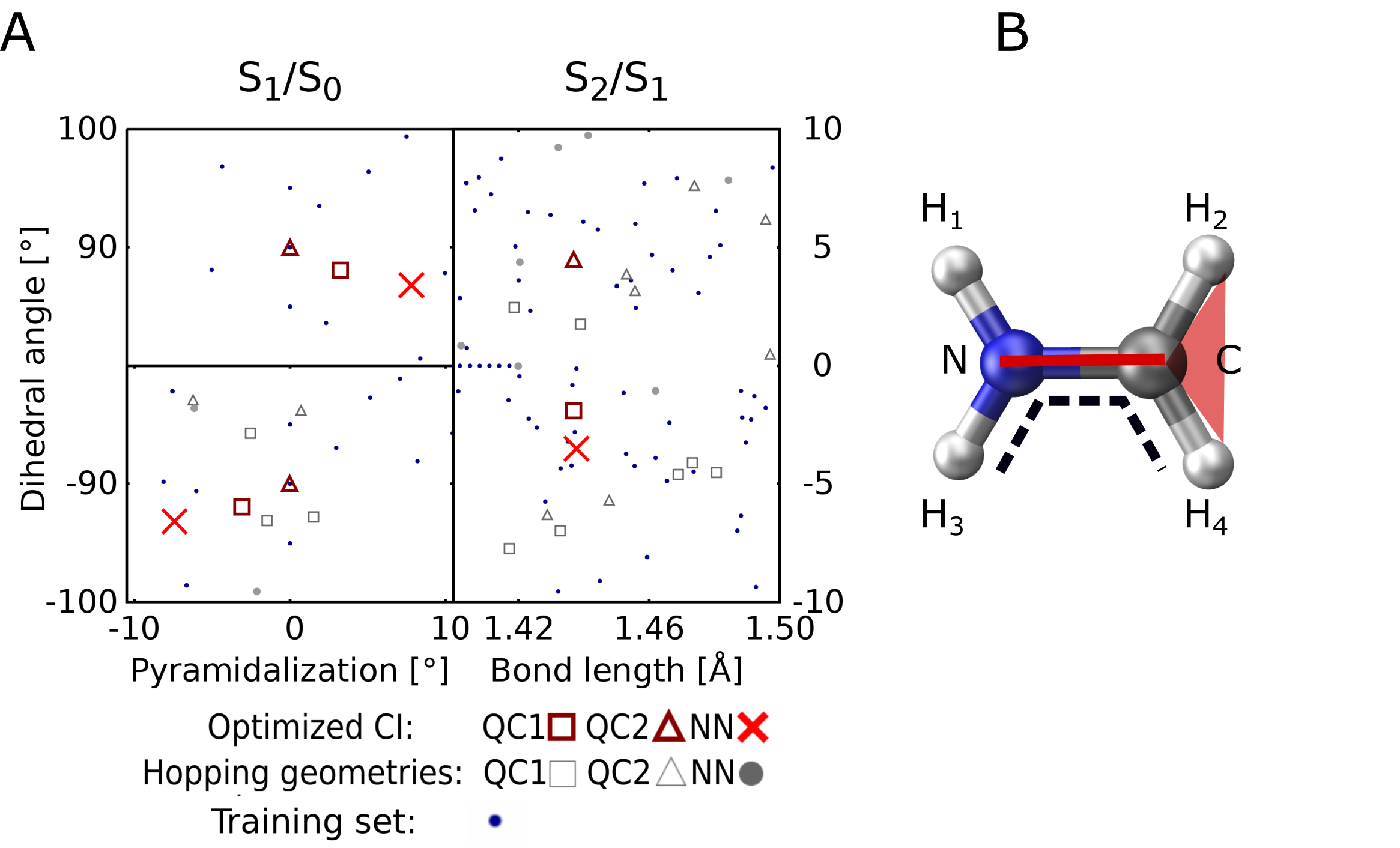}
    \caption{\textbf{Conical intersection analysis in addition to Fig. 5 in the main text.} \textbf{(A)} Scatter plots showing the distribution of hopping geometries of each method as well as optimized nuclear configurations close to the minimum energy CI present between the S$_1$ state and the ground state, S$_0$, (left) as well as between the S$_2$ state and the S$_1$ state (right). Geometries from the training set with 4000 data points are plotted as blue circles.} \textbf{(B)} Shown is the equilibrium geometry of methylenimmonium cation in order to specify the atoms that describe the used dihedral and pyramidalization angle in Fig. 5 and Fig. 6 in the main text and Fig. \ref{fig:CIgeom} here. The pyramidalization angle that is used to describe the S$_1$/S$_0$ CI is defined by the angle between the bond between the nitrogen (N) and the carbon (C) atom (red line) and the plane that is spanned by the C atom and the two hydrogen atoms, H2 and H4 (red triangle). The angle between the planes spanned by atoms H3-N-C and N-C-H4 is chosen to distinguish geometries describing both CIs is defined by hydrogen atoms, H3 and H4, as well as by the C and N atom (dashed black line).
    \label{fig:CI}
\end{figure}

The S$_1$/S$_0$ CI shows a rotation of the molecule around the H3-N-C-H4 dihedral angle (see Fig. \ref{fig:CI} for the definition) of $\sim90^{\circ}$. Due to symmetry of the molecule, the CI is obtained at $\sim90^{\circ}$ as well as at $\sim-90^{\circ}$. The optimized geometries agree well between all methods (\textit{QC1}, \textit{QC2}, NN). All methods yield hopping geometries that are distributed in a rather large region around these points. The S$_2$/S$_1$ CI, is defined by a slight bi-pyramidalization and a bond elongation between the carbon and the nitrogen atom of around 1.44 \AA ~(see Fig. \ref{fig:CIgeom}) compared to a bond length of around 1.29 \AA ~at the equilibrium geometry (Fig. \ref{fig:CI}B). Again, molecular geometries obtained with different methods are very similar. The scatter plot in the right panel of Fig. \ref{fig:CI}A shows the good agreement among different methods and also depicts the distribution of data points in the final training set. 

\begin{table}[h]
\centering
\small
  \caption{Cartesian coordinates of the S$_1$/S$_0$ CI obtained from deep NNs. }
 \label{tab:CifirstNN}
\begin{tabular}{llll}
\hline
\hline
Atom &x-coordinate [\AA] &y-coordinate [\AA] &z-coordinate [\AA]\\
\hline
C  & -0.09510	&0.0547	&-0.1529\\
N &0.0928	&0.0520	&1.2196\\
H &0.6760	&0.7307	&-0.6251\\
H &0.7899	&-0.6147	&1.8341\\
H &-0.5741	&0.8014	&1.7538\\
H &-0.7677	&-0.5838	&-0.7111\\
\hline
\end{tabular}
\centering
\end{table}

\begin{table}[h]
\centering
\small
  \caption{Cartesian coordinates of the S$_1$/S$_0$ CI obtained from MR-CISD/aug-cc-pVDZ. }
 \label{tab:CifirstQC}
\begin{tabular}{llll}
\hline
\hline
Atom &x-coordinate [\AA] &y-coordinate [\AA] &z-coordinate [\AA]\\
\hline
C	&-0.067495	&-0.271367	&-0.083550\\
N	&0.002414	&-0.191463	&1.316026\\
H	&-0.264937	&-1.250001	&-0.530777\\
H	&0.882335	&-0.347469	&1.825166\\
H	&-0.826010	&0.016867	&1.888644\\
H	&0.133036	&0.637597	&-0.658113\\
\hline
\end{tabular}
\centering
\end{table}

\begin{table}[h]
\centering
\small
  \caption{Cartesian coordinates of the S$_2$/S$_1$ CI obtained from deep NNs. }
 \label{tab:CilastNN}
\begin{tabular}{llll}
\hline
\hline
Atom &x-coordinate [\AA] &y-coordinate [\AA] &z-coordinate [\AA]\\
\hline
C	&0.092590	&0.05507	&-0.03636\\
N	&-0.02183	&0.01101	&1.3961\\
H	&0.9094	&-0.2682	&-0.6655\\
H	&0.6959	&-0.4079	&1.9997\\
H	&-0.7125	&-0.46625	&1.9366\\
H	&-0.8898	&-0.3630	&-0.4760\\
\hline
\end{tabular}
\centering
\end{table}

\begin{table}[h]
\centering
\small
  \caption{Cartesian coordinates of the S$_2$/S$_1$ CI obtained from MR-CISD/aug-cc-pVDZ. }
 \label{tab:CilastQC}
\begin{tabular}{llll}
\hline
\hline
Atom &x-coordinate [\AA] &y-coordinate [\AA] &z-coordinate [\AA]\\
\hline
C	&0.049953	&-0.345208	&-0.016365\\
N	&0.033847	&-0.254299	&1.417459\\
H	&0.910037	&0.088100	&-0.642754\\
H	&0.869788	&0.200682	&1.810525\\
H	&-0.809643	&0.202872	&1.791428\\
H	&-0.793958	&0.090807	&-0.662521\\
\hline
\end{tabular}
\centering
\end{table}



\clearpage
\section{Future perspectives}
As an outlook we expect our approach to offer new possibilities to study also excited-state dynamics of larger molecules on longer timescales. By using the adaptive sampling procedure in combination with NN dynamics simulations, an initial training set can be automatically expanded until the conformational space of a molecule important for dynamics is sampled comprehensively. In the following, a short discussion on expected scaling of the methods and possibilities to treat larger molecules is given.

As a good starting point for the initial training set we suggest sampling of normal modes and linear combinations of different normal modes. Alternatively, the points obtained from a geometry optimization of critical points like CIs might deliver an even better initial set. As a guideline, we suggest from our limited experience that approximately 1000 data points should be considered for the initial training set. With this procedure, some critical regions of the potential energy surface close to the equilibrium geometry can already be covered in the training set. Further sampling of reaction coordinates that are considered to be important during excited-state molecular dynamics simulations seems to be a good guess. As an example, a reaction coordinate that involves the dissociation of an atom, can be included. 
Dependent on the size and complexity of the molecule as well as on the degree of initial sampling, the trajectories will be interrupted more frequently in the initial stages of an adaptive sampling run. In terms of training set sizes in general, we do not expect a linear scaling along with the number of atoms. The number of necessary data points does not only depend on the number of atoms of a molecule, but also on its excited state dynamics dictated for example by its flexibility or the excitation energy. The more processes that can take place after excitation of a molecule, the larger the space is that needs to be covered. Further, the number of states included in dynamics simulations as well as the quantum chemical reference method play a role and should be considered. In general, no rule can be given in terms of optimal training set size, but the convergence behavior of the adaptive sampling runs can serve as an indicator on whether additional reference data is required. Finally, we want to point out that using the adaptive sampling scheme -- as soon as enough data points for a machine learning model can be obtained within a reasonable time frame -- seems to be most efficient for sampling the relevant space visited during a dynamics simulation.
We further recommend to take special care when selecting the number of electronic states. As already discussed in the main text within the section on the phase correction algorithm, it is important to compute more electronic states for molecules that have a lot of electronic states lying close in energy. However, those additional states should only be used for phase correction and computation of wavefunction overlaps and do not need to be considered for couplings or computation of forces. With this approach, problems arising due to "intruder states" during the phase correction procedure can be avoided.
Regarding the network architecture, we recommend to adapt it to the complexity of the molecule and expect approximately linear scaling of both, the size of the network and its computational performance, with the number of atoms of a molecule.

Last but not least, we want to comment on non-molecule specific excited-state potentials, similar to the idea of universal force fields that are in part already available for the ground state. We expect this to be a very complex task that has to be considered with a lot of care. If molecules show similar excited state dynamics -- for example the methylenimmonium cation and the ethylene molecule~\cite{Sellner2013MP} are isoelectronic and show a rotation along the dihedral angle for the S$_1$/S$_0$ transition -- a non-molecule specific method can be thought of to describe this transition. However, the order of higher-lying states is different in these two isoelectronic species and higher excitation energies will induce completely different dynamics in the respective systems. Therefore and also due to the problematic generalization and molecule-specific couplings and excited-state potentials, we do not expect a non-molecule specific model to be available in the near future.


\begin{mcitethebibliography}{85}
\providecommand*{\natexlab}[1]{#1}
\providecommand*{\mciteSetBstSublistMode}[1]{}
\providecommand*{\mciteSetBstMaxWidthForm}[2]{}
\providecommand*{\mciteBstWouldAddEndPuncttrue}
  {\def\EndOfBibitem{\unskip.}}
\providecommand*{\mciteBstWouldAddEndPunctfalse}
  {\let\EndOfBibitem\relax}
\providecommand*{\mciteSetBstMidEndSepPunct}[3]{}
\providecommand*{\mciteSetBstSublistLabelBeginEnd}[3]{}
\providecommand*{\EndOfBibitem}{}
\mciteSetBstSublistMode{f}
\mciteSetBstMaxWidthForm{subitem}
{(\emph{\alph{mcitesubitemcount}})}
\mciteSetBstSublistLabelBeginEnd{\mcitemaxwidthsubitemform\space}
{\relax}{\relax}

\bibitem[Goodfellow \emph{et~al.}(2016)Goodfellow, Bengio, and
  Courville]{Goodfellow2016}
I.~Goodfellow, Y.~Bengio and A.~Courville, \emph{Deep Learning}, MIT Press,
  2016\relax
\mciteBstWouldAddEndPuncttrue
\mciteSetBstMidEndSepPunct{\mcitedefaultmidpunct}
{\mcitedefaultendpunct}{\mcitedefaultseppunct}\relax
\EndOfBibitem
\bibitem[Silver \emph{et~al.}(2016)Silver, Huang, Maddison, Guez, Sifre,
  van~den Driessche, Schrittwieser, Antonoglou, Panneershelvam, Lanctot,
  Dieleman, Grewe, Nham, Kalchbrenner, Sutskever, Lillicrap, Leach,
  Kavukcuoglu, Graepel, and Hassabis]{Silver2016N}
D.~Silver, A.~Huang, C.~J. Maddison, A.~Guez, L.~Sifre, G.~van~den Driessche,
  J.~Schrittwieser, I.~Antonoglou, V.~Panneershelvam, M.~Lanctot, S.~Dieleman,
  D.~Grewe, J.~Nham, N.~Kalchbrenner, I.~Sutskever, T.~Lillicrap, M.~Leach,
  K.~Kavukcuoglu, T.~Graepel and D.~Hassabis, \emph{Nature}, 2016,
  \textbf{529}, 484--489\relax
\mciteBstWouldAddEndPuncttrue
\mciteSetBstMidEndSepPunct{\mcitedefaultmidpunct}
{\mcitedefaultendpunct}{\mcitedefaultseppunct}\relax
\EndOfBibitem
\bibitem[Bansak \emph{et~al.}(2018)Bansak, Ferwerda, Hainmueller, Dillon,
  Hangartner, Lawrence, and Weinstein]{Bansak2018S}
K.~Bansak, J.~Ferwerda, J.~Hainmueller, A.~Dillon, D.~Hangartner, D.~Lawrence
  and J.~Weinstein, \emph{Science}, 2018, \textbf{359}, 325--329\relax
\mciteBstWouldAddEndPuncttrue
\mciteSetBstMidEndSepPunct{\mcitedefaultmidpunct}
{\mcitedefaultendpunct}{\mcitedefaultseppunct}\relax
\EndOfBibitem
\bibitem[Sanchez-Lengeling and Aspuru-Guzik(2018)]{Sanchez-Lengeling2018S}
B.~Sanchez-Lengeling and A.~Aspuru-Guzik, \emph{Science}, 2018, \textbf{361},
  360--365\relax
\mciteBstWouldAddEndPuncttrue
\mciteSetBstMidEndSepPunct{\mcitedefaultmidpunct}
{\mcitedefaultendpunct}{\mcitedefaultseppunct}\relax
\EndOfBibitem
\bibitem[Butler \emph{et~al.}(2018)Butler, Davies, Cartwright, Isayev, and
  Walsh]{Butler2018N}
K.~T. Butler, D.~W. Davies, H.~Cartwright, O.~Isayev and A.~Walsh,
  \emph{Nature}, 2018, \textbf{559}, 547--555\relax
\mciteBstWouldAddEndPuncttrue
\mciteSetBstMidEndSepPunct{\mcitedefaultmidpunct}
{\mcitedefaultendpunct}{\mcitedefaultseppunct}\relax
\EndOfBibitem
\bibitem[Goldsmith \emph{et~al.}(2018)Goldsmith, Esterhuizen, Liu, Bartel, and
  Sutton]{Goldsmith2018AJ}
B.~R. Goldsmith, J.~Esterhuizen, J.-X. Liu, C.~J. Bartel and C.~Sutton,
  \emph{AIChE J.}, 2018, \textbf{64}, 2311--2323\relax
\mciteBstWouldAddEndPuncttrue
\mciteSetBstMidEndSepPunct{\mcitedefaultmidpunct}
{\mcitedefaultendpunct}{\mcitedefaultseppunct}\relax
\EndOfBibitem
\bibitem[Cerullo \emph{et~al.}(2002)Cerullo, Polli, Lanzani, De~Silvestri,
  Hashimoto, and Cogdell]{Cerullo2002S}
G.~Cerullo, D.~Polli, G.~Lanzani, S.~De~Silvestri, H.~Hashimoto and R.~J.
  Cogdell, \emph{Science}, 2002, \textbf{298}, 2395--2398\relax
\mciteBstWouldAddEndPuncttrue
\mciteSetBstMidEndSepPunct{\mcitedefaultmidpunct}
{\mcitedefaultendpunct}{\mcitedefaultseppunct}\relax
\EndOfBibitem
\bibitem[Schultz \emph{et~al.}(2004)Schultz, Samoylova, Radloff, Hertel,
  Sobolewski, and Domcke]{Schultz2004S}
T.~Schultz, E.~Samoylova, W.~Radloff, I.~V. Hertel, A.~L. Sobolewski and
  W.~Domcke, \emph{Science}, 2004, \textbf{306}, 1765--1768\relax
\mciteBstWouldAddEndPuncttrue
\mciteSetBstMidEndSepPunct{\mcitedefaultmidpunct}
{\mcitedefaultendpunct}{\mcitedefaultseppunct}\relax
\EndOfBibitem
\bibitem[Schreier \emph{et~al.}(2007)Schreier, Schrader, Koller, Gilch,
  Crespo-Hern\'{a}ndez, Swaminathan, Charell, Zinth, and Kohler]{Schreier2007S}
W.~J. Schreier, T.~E. Schrader, F.~O. Koller, P.~Gilch, C.~E.
  Crespo-Hern\'{a}ndez, V.~N. Swaminathan, T.~Charell, W.~Zinth and B.~Kohler,
  \emph{Science}, 2007, \textbf{315}, 625--629\relax
\mciteBstWouldAddEndPuncttrue
\mciteSetBstMidEndSepPunct{\mcitedefaultmidpunct}
{\mcitedefaultendpunct}{\mcitedefaultseppunct}\relax
\EndOfBibitem
\bibitem[Rauer \emph{et~al.}(2016)Rauer, Nogueira, Marquetand, and
  Gonz\'alez]{Rauer2016JACS}
C.~Rauer, J.~J. Nogueira, P.~Marquetand and L.~Gonz\'alez, \emph{J. Am. Chem.
  Soc.}, 2016, \textbf{138}, 15911--15916\relax
\mciteBstWouldAddEndPuncttrue
\mciteSetBstMidEndSepPunct{\mcitedefaultmidpunct}
{\mcitedefaultendpunct}{\mcitedefaultseppunct}\relax
\EndOfBibitem
\bibitem[Romero \emph{et~al.}(2017)Romero, Novoderezhkin, and
  Grondelle]{Romero2017N}
E.~Romero, V.~I. Novoderezhkin and R.~v. Grondelle, \emph{Nature}, 2017,
  \textbf{543}, 355\relax
\mciteBstWouldAddEndPuncttrue
\mciteSetBstMidEndSepPunct{\mcitedefaultmidpunct}
{\mcitedefaultendpunct}{\mcitedefaultseppunct}\relax
\EndOfBibitem
\bibitem[Ahmad \emph{et~al.}(2016)Ahmad, Ahmed, Anwar, Sheraz, and
  Sikorski]{Ahmad2016IJP}
I.~Ahmad, S.~Ahmed, Z.~Anwar, M.~A. Sheraz and M.~Sikorski, \emph{Int. J.
  Photoenergy}, 2016, \textbf{2016}, 1--19\relax
\mciteBstWouldAddEndPuncttrue
\mciteSetBstMidEndSepPunct{\mcitedefaultmidpunct}
{\mcitedefaultendpunct}{\mcitedefaultseppunct}\relax
\EndOfBibitem
\bibitem[Mathew \emph{et~al.}(2014)Mathew, Yella, Gao, Humphry-Baker, Curchod,
  Ashari-Astani, Tavernelli, Rothlisberger, Nazeeruddin, and
  Gr\"{a}tzel]{Mathew2014NC}
S.~Mathew, A.~Yella, P.~Gao, R.~Humphry-Baker, B.~F.~E. Curchod,
  N.~Ashari-Astani, I.~Tavernelli, U.~Rothlisberger, M.~K. Nazeeruddin and
  M.~Gr\"{a}tzel, \emph{Nat. Chem.}, 2014, \textbf{6}, 242\relax
\mciteBstWouldAddEndPuncttrue
\mciteSetBstMidEndSepPunct{\mcitedefaultmidpunct}
{\mcitedefaultendpunct}{\mcitedefaultseppunct}\relax
\EndOfBibitem
\bibitem[Bart{\'o}k \emph{et~al.}(2017)Bart{\'o}k, De, Poelking, Bernstein,
  Kermode, Cs{\'a}nyi, and Ceriotti]{Bartok2017SA}
A.~P. Bart{\'o}k, S.~De, C.~Poelking, N.~Bernstein, J.~R. Kermode,
  G.~Cs{\'a}nyi and M.~Ceriotti, \emph{Sci. Adv.}, 2017, \textbf{3},
  e1701816\relax
\mciteBstWouldAddEndPuncttrue
\mciteSetBstMidEndSepPunct{\mcitedefaultmidpunct}
{\mcitedefaultendpunct}{\mcitedefaultseppunct}\relax
\EndOfBibitem
\bibitem[Mai \emph{et~al.}(2018)Mai, Marquetand, and Gonz\'alez]{Mai2018WCMS}
S.~Mai, P.~Marquetand and L.~Gonz\'alez, \emph{WIREs Comput. Mol. Sci.}, 2018,
  \textbf{8}, e1370\relax
\mciteBstWouldAddEndPuncttrue
\mciteSetBstMidEndSepPunct{\mcitedefaultmidpunct}
{\mcitedefaultendpunct}{\mcitedefaultseppunct}\relax
\EndOfBibitem
\bibitem[Chmiela \emph{et~al.}(2018)Chmiela, Sauceda, M\"{u}ller, and
  Tkatchenko]{Chmiela2018NC}
S.~Chmiela, H.~E. Sauceda, K.-R. M\"{u}ller and A.~Tkatchenko, \emph{Nat.
  Commun.}, 2018, \textbf{9}, 3887\relax
\mciteBstWouldAddEndPuncttrue
\mciteSetBstMidEndSepPunct{\mcitedefaultmidpunct}
{\mcitedefaultendpunct}{\mcitedefaultseppunct}\relax
\EndOfBibitem
\bibitem[Rupp(2015)]{Rupp2015IJQC}
M.~Rupp, \emph{Int. J. Quantum Chem.}, 2015, \textbf{115}, 1058--1073\relax
\mciteBstWouldAddEndPuncttrue
\mciteSetBstMidEndSepPunct{\mcitedefaultmidpunct}
{\mcitedefaultendpunct}{\mcitedefaultseppunct}\relax
\EndOfBibitem
\bibitem[{von Lilienfeld}(2018)]{Lilienfeld2018ACIE}
O.~A. {von Lilienfeld}, \emph{Angew. Chem. Int. Ed.}, 2018, \textbf{57},
  4164--4169\relax
\mciteBstWouldAddEndPuncttrue
\mciteSetBstMidEndSepPunct{\mcitedefaultmidpunct}
{\mcitedefaultendpunct}{\mcitedefaultseppunct}\relax
\EndOfBibitem
\bibitem[Bart\'ok \emph{et~al.}(2010)Bart\'ok, Payne, Kondor, and
  Cs\'anyi]{Bartok2010PRL}
A.~P. Bart\'ok, M.~C. Payne, R.~Kondor and G.~Cs\'anyi, \emph{Phys. Rev.
  Lett.}, 2010, \textbf{104}, 136403\relax
\mciteBstWouldAddEndPuncttrue
\mciteSetBstMidEndSepPunct{\mcitedefaultmidpunct}
{\mcitedefaultendpunct}{\mcitedefaultseppunct}\relax
\EndOfBibitem
\bibitem[Li \emph{et~al.}(2015)Li, Kermode, and De~Vita]{Li2015PRL}
Z.~Li, J.~R. Kermode and A.~De~Vita, \emph{Phys. Rev. Lett.}, 2015,
  \textbf{114}, 096405\relax
\mciteBstWouldAddEndPuncttrue
\mciteSetBstMidEndSepPunct{\mcitedefaultmidpunct}
{\mcitedefaultendpunct}{\mcitedefaultseppunct}\relax
\EndOfBibitem
\bibitem[Rupp \emph{et~al.}(2015)Rupp, Ramakrishnan, and von
  Lilienfeld]{Rupp2015JPCL}
M.~Rupp, R.~Ramakrishnan and O.~A. von Lilienfeld, \emph{J. Phys. Chem. Lett.},
  2015, \textbf{6}, 3309--3313\relax
\mciteBstWouldAddEndPuncttrue
\mciteSetBstMidEndSepPunct{\mcitedefaultmidpunct}
{\mcitedefaultendpunct}{\mcitedefaultseppunct}\relax
\EndOfBibitem
\bibitem[Behler(2016)]{Behler2016JCP}
J.~Behler, \emph{J. Chem. Phys.}, 2016, \textbf{145}, 170901\relax
\mciteBstWouldAddEndPuncttrue
\mciteSetBstMidEndSepPunct{\mcitedefaultmidpunct}
{\mcitedefaultendpunct}{\mcitedefaultseppunct}\relax
\EndOfBibitem
\bibitem[Gastegger \emph{et~al.}(2017)Gastegger, Behler, and
  Marquetand]{Gastegger2017CS}
M.~Gastegger, J.~Behler and P.~Marquetand, \emph{Chem. Sci.}, 2017, \textbf{8},
  6924--6935\relax
\mciteBstWouldAddEndPuncttrue
\mciteSetBstMidEndSepPunct{\mcitedefaultmidpunct}
{\mcitedefaultendpunct}{\mcitedefaultseppunct}\relax
\EndOfBibitem
\bibitem[Botu \emph{et~al.}(2017)Botu, Batra, Chapman, and
  Ramprasad]{Botu2017JPCC}
V.~Botu, R.~Batra, J.~Chapman and R.~Ramprasad, \emph{J. Phys. Chem. C}, 2017,
  \textbf{121}, 511--522\relax
\mciteBstWouldAddEndPuncttrue
\mciteSetBstMidEndSepPunct{\mcitedefaultmidpunct}
{\mcitedefaultendpunct}{\mcitedefaultseppunct}\relax
\EndOfBibitem
\bibitem[Smith \emph{et~al.}(2017)Smith, Isayev, and Roitberg]{Smith2017CS}
J.~S. Smith, O.~Isayev and A.~E. Roitberg, \emph{Chem. Sci.}, 2017, \textbf{8},
  3192--3203\relax
\mciteBstWouldAddEndPuncttrue
\mciteSetBstMidEndSepPunct{\mcitedefaultmidpunct}
{\mcitedefaultendpunct}{\mcitedefaultseppunct}\relax
\EndOfBibitem
\bibitem[Behler(2017)]{Behler2017ACIE}
J.~Behler, \emph{Angew. Chem. Int. Edit.}, 2017, \textbf{56},
  12828--12840\relax
\mciteBstWouldAddEndPuncttrue
\mciteSetBstMidEndSepPunct{\mcitedefaultmidpunct}
{\mcitedefaultendpunct}{\mcitedefaultseppunct}\relax
\EndOfBibitem
\bibitem[Zong \emph{et~al.}(2018)Zong, Pilania, Ding, Ackland, and
  Lookman]{Zong2018npjCM}
H.~Zong, G.~Pilania, X.~Ding, G.~J. Ackland and T.~Lookman, \emph{npj Comput.
  Mater.}, 2018, \textbf{4}, 48\relax
\mciteBstWouldAddEndPuncttrue
\mciteSetBstMidEndSepPunct{\mcitedefaultmidpunct}
{\mcitedefaultendpunct}{\mcitedefaultseppunct}\relax
\EndOfBibitem
\bibitem[Bart\'ok \emph{et~al.}(2018)Bart\'ok, Kermode, Bernstein, and
  Cs\'anyi]{Bartok2018PRX}
A.~P. Bart\'ok, J.~Kermode, N.~Bernstein and G.~Cs\'anyi, \emph{Phys. Rev. X},
  2018, \textbf{8}, 041048\relax
\mciteBstWouldAddEndPuncttrue
\mciteSetBstMidEndSepPunct{\mcitedefaultmidpunct}
{\mcitedefaultendpunct}{\mcitedefaultseppunct}\relax
\EndOfBibitem
\bibitem[Xia and Kais(2018)]{Xia2018NC}
R.~Xia and S.~Kais, \emph{Nat. Commun.}, 2018, \textbf{9}, 4195\relax
\mciteBstWouldAddEndPuncttrue
\mciteSetBstMidEndSepPunct{\mcitedefaultmidpunct}
{\mcitedefaultendpunct}{\mcitedefaultseppunct}\relax
\EndOfBibitem
\bibitem[Chan \emph{et~al.}(2019)Chan, Narayanan, Cherukara, Sen, Sasikumar,
  Gray, Chan, and Sankaranarayanan]{Chan2019JPCC}
H.~Chan, B.~Narayanan, M.~J. Cherukara, F.~G. Sen, K.~Sasikumar, S.~K. Gray,
  M.~K.~Y. Chan and S.~K. R.~S. Sankaranarayanan, \emph{J. Phys. Chem. C},
  2019, \textbf{123}, 6941\relax
\mciteBstWouldAddEndPuncttrue
\mciteSetBstMidEndSepPunct{\mcitedefaultmidpunct}
{\mcitedefaultendpunct}{\mcitedefaultseppunct}\relax
\EndOfBibitem
\bibitem[Christensen \emph{et~al.}(2019)Christensen, Faber, and von
  Lilienfeld]{Christensen2019JCP}
A.~S. Christensen, F.~A. Faber and O.~A. von Lilienfeld, \emph{J. Chem. Phys.},
  2019, \textbf{150}, 064105\relax
\mciteBstWouldAddEndPuncttrue
\mciteSetBstMidEndSepPunct{\mcitedefaultmidpunct}
{\mcitedefaultendpunct}{\mcitedefaultseppunct}\relax
\EndOfBibitem
\bibitem[Netzloff \emph{et~al.}(2006)Netzloff, Collins, and
  Gordon]{Netzloff2006JCP}
H.~M. Netzloff, M.~A. Collins and M.~S. Gordon, \emph{J. Chem. Phys.}, 2006,
  \textbf{124}, 154104\relax
\mciteBstWouldAddEndPuncttrue
\mciteSetBstMidEndSepPunct{\mcitedefaultmidpunct}
{\mcitedefaultendpunct}{\mcitedefaultseppunct}\relax
\EndOfBibitem
\bibitem[Bettens and Collins(1999)]{Bettens1999JCP}
R.~P.~A. Bettens and M.~A. Collins, \emph{J. Chem. Phys.}, 1999, \textbf{111},
  816--826\relax
\mciteBstWouldAddEndPuncttrue
\mciteSetBstMidEndSepPunct{\mcitedefaultmidpunct}
{\mcitedefaultendpunct}{\mcitedefaultseppunct}\relax
\EndOfBibitem
\bibitem[Behler \emph{et~al.}(2008)Behler, Reuter, and
  Scheffler]{Behler2008PRB}
J.~Behler, K.~Reuter and M.~Scheffler, \emph{Phys. Rev. B}, 2008, \textbf{77},
  115421\relax
\mciteBstWouldAddEndPuncttrue
\mciteSetBstMidEndSepPunct{\mcitedefaultmidpunct}
{\mcitedefaultendpunct}{\mcitedefaultseppunct}\relax
\EndOfBibitem
\bibitem[Carbogno \emph{et~al.}(2010)Carbogno, Behler, Reuter, and
  Gro\ss{}]{Carbogno2010PRB}
C.~Carbogno, J.~Behler, K.~Reuter and A.~Gro\ss{}, \emph{Phys. Rev. B}, 2010,
  \textbf{81}, 035410\relax
\mciteBstWouldAddEndPuncttrue
\mciteSetBstMidEndSepPunct{\mcitedefaultmidpunct}
{\mcitedefaultendpunct}{\mcitedefaultseppunct}\relax
\EndOfBibitem
\bibitem[H\"ase \emph{et~al.}(2016)H\"ase, Valleau, Pyzer-Knapp, and
  Aspuru-Guzik]{Haese2016CS}
F.~H\"ase, S.~Valleau, E.~Pyzer-Knapp and A.~Aspuru-Guzik, \emph{Chem. Sci.},
  2016, \textbf{7}, 5139--5147\relax
\mciteBstWouldAddEndPuncttrue
\mciteSetBstMidEndSepPunct{\mcitedefaultmidpunct}
{\mcitedefaultendpunct}{\mcitedefaultseppunct}\relax
\EndOfBibitem
\bibitem[Liu \emph{et~al.}(2017)Liu, Du, Zhang, and Gao]{Liu2017SR}
F.~Liu, L.~Du, D.~Zhang and J.~Gao, \emph{Sci. Rep.}, 2017, \textbf{7},
  1--12\relax
\mciteBstWouldAddEndPuncttrue
\mciteSetBstMidEndSepPunct{\mcitedefaultmidpunct}
{\mcitedefaultendpunct}{\mcitedefaultseppunct}\relax
\EndOfBibitem
\bibitem[Hu \emph{et~al.}(2018)Hu, Xie, Li, Li, and Lan]{Hu2018JPCL}
D.~Hu, Y.~Xie, X.~Li, L.~Li and Z.~Lan, \emph{J. Phys. Chem. Lett.}, 2018,
  \textbf{9}, 2725--2732\relax
\mciteBstWouldAddEndPuncttrue
\mciteSetBstMidEndSepPunct{\mcitedefaultmidpunct}
{\mcitedefaultendpunct}{\mcitedefaultseppunct}\relax
\EndOfBibitem
\bibitem[Dral \emph{et~al.}(2018)Dral, Barbatti, and Thiel]{Dral2018JPCL}
P.~O. Dral, M.~Barbatti and W.~Thiel, \emph{J. Phys. Chem. Lett.}, 2018,
  \textbf{9}, 5660--5663\relax
\mciteBstWouldAddEndPuncttrue
\mciteSetBstMidEndSepPunct{\mcitedefaultmidpunct}
{\mcitedefaultendpunct}{\mcitedefaultseppunct}\relax
\EndOfBibitem
\bibitem[Chen \emph{et~al.}(2018)Chen, Liu, Fang, Dral, and Cui]{Chen2018JPCL}
W.-K. Chen, X.-Y. Liu, W.-H. Fang, P.~O. Dral and G.~Cui, \emph{J. Phys. Chem.
  Lett.}, 2018, \textbf{9}, 6702--6708\relax
\mciteBstWouldAddEndPuncttrue
\mciteSetBstMidEndSepPunct{\mcitedefaultmidpunct}
{\mcitedefaultendpunct}{\mcitedefaultseppunct}\relax
\EndOfBibitem
\bibitem[Williams and Eisfeld(2018)]{William2018JCP}
D.~M.~G. Williams and W.~Eisfeld, \emph{J. Chem. Phys.}, 2018, \textbf{149},
  204106\relax
\mciteBstWouldAddEndPuncttrue
\mciteSetBstMidEndSepPunct{\mcitedefaultmidpunct}
{\mcitedefaultendpunct}{\mcitedefaultseppunct}\relax
\EndOfBibitem
\bibitem[Xie \emph{et~al.}(2018)Xie, Zhu, Yarkony, and Guo]{Xie2018JCP}
C.~Xie, X.~Zhu, D.~R. Yarkony and H.~Guo, \emph{J. Chem. Phys.}, 2018,
  \textbf{149}, 144107\relax
\mciteBstWouldAddEndPuncttrue
\mciteSetBstMidEndSepPunct{\mcitedefaultmidpunct}
{\mcitedefaultendpunct}{\mcitedefaultseppunct}\relax
\EndOfBibitem
\bibitem[Guan \emph{et~al.}(2019)Guan, Zhang, Guo, and Yarkony]{Guan2019PCCP}
Y.~Guan, D.~H. Zhang, H.~Guo and D.~R. Yarkony, \emph{Phys. Chem. Chem. Phys.},
  2019,  DOI:10.1039/C8CP06598E\relax
\mciteBstWouldAddEndPuncttrue
\mciteSetBstMidEndSepPunct{\mcitedefaultmidpunct}
{\mcitedefaultendpunct}{\mcitedefaultseppunct}\relax
\EndOfBibitem
\bibitem[Domcke \emph{et~al.}(2004)Domcke, Yarkony, and K\"{o}ppel]{Domcke2004}
W.~Domcke, D.~R. Yarkony and H.~K\"{o}ppel, \emph{Conical Intersections:
  Electronic Structure, Dynamics and Spectroscopy}, WORLD SCIENTIFIC,
  2004\relax
\mciteBstWouldAddEndPuncttrue
\mciteSetBstMidEndSepPunct{\mcitedefaultmidpunct}
{\mcitedefaultendpunct}{\mcitedefaultseppunct}\relax
\EndOfBibitem
\bibitem[Richter \emph{et~al.}(2011)Richter, Marquetand,
  Gonz\'{a}lez-V\'{a}zquez, Sola, and Gonz\'{a}lez]{Richter2011JCTC}
M.~Richter, P.~Marquetand, J.~Gonz\'{a}lez-V\'{a}zquez, I.~Sola and
  L.~Gonz\'{a}lez, \emph{J. Chem. Theory Comput.}, 2011, \textbf{7},
  1253--1258\relax
\mciteBstWouldAddEndPuncttrue
\mciteSetBstMidEndSepPunct{\mcitedefaultmidpunct}
{\mcitedefaultendpunct}{\mcitedefaultseppunct}\relax
\EndOfBibitem
\bibitem[Tully(1990)]{Tully1990JCP}
J.~C. Tully, \emph{J. Chem. Phys.}, 1990, \textbf{93}, 1061--1071\relax
\mciteBstWouldAddEndPuncttrue
\mciteSetBstMidEndSepPunct{\mcitedefaultmidpunct}
{\mcitedefaultendpunct}{\mcitedefaultseppunct}\relax
\EndOfBibitem
\bibitem[Mai \emph{et~al.}(2018)Mai, Richter, Ruckenbauer, Oppel, Marquetand,
  and Gonz\'alez]{sharc-md2}
S.~Mai, M.~Richter, M.~Ruckenbauer, M.~Oppel, P.~Marquetand and L.~Gonz\'alez,
  \emph{SHARC2.0: Surface Hopping Including Arbitrary Couplings -- Program
  Package for Non-Adiabatic Dynamics}, sharc-md.org, 2018\relax
\mciteBstWouldAddEndPuncttrue
\mciteSetBstMidEndSepPunct{\mcitedefaultmidpunct}
{\mcitedefaultendpunct}{\mcitedefaultseppunct}\relax
\EndOfBibitem
\bibitem[Behler(2015)]{Behler2015IJQC}
J.~Behler, \emph{Int. J. Quantum Chem.}, 2015, \textbf{115}, 1032--1050\relax
\mciteBstWouldAddEndPuncttrue
\mciteSetBstMidEndSepPunct{\mcitedefaultmidpunct}
{\mcitedefaultendpunct}{\mcitedefaultseppunct}\relax
\EndOfBibitem
\bibitem[Botu and Ramprasad(2015)]{Botu2015IJQC}
V.~Botu and R.~Ramprasad, \emph{Int. J. Quantum Chem.}, 2015, \textbf{115},
  1074--1083\relax
\mciteBstWouldAddEndPuncttrue
\mciteSetBstMidEndSepPunct{\mcitedefaultmidpunct}
{\mcitedefaultendpunct}{\mcitedefaultseppunct}\relax
\EndOfBibitem
\bibitem[Smith \emph{et~al.}(2018)Smith, Nebgen, Lubbers, Isayev, and
  Roitberg]{Smith2018JCP}
J.~S. Smith, B.~Nebgen, N.~Lubbers, O.~Isayev and A.~E. Roitberg, \emph{J.
  Chem. Phys.}, 2018, \textbf{148}, 241733\relax
\mciteBstWouldAddEndPuncttrue
\mciteSetBstMidEndSepPunct{\mcitedefaultmidpunct}
{\mcitedefaultendpunct}{\mcitedefaultseppunct}\relax
\EndOfBibitem
\bibitem[Marquetand \emph{et~al.}(2004)Marquetand, Materny, Henriksen, and
  Engel]{Marquetand2004JCP}
P.~Marquetand, A.~Materny, N.~E. Henriksen and V.~Engel, \emph{J. Chem. Phys.},
  2004, \textbf{120}, 5871--5874\relax
\mciteBstWouldAddEndPuncttrue
\mciteSetBstMidEndSepPunct{\mcitedefaultmidpunct}
{\mcitedefaultendpunct}{\mcitedefaultseppunct}\relax
\EndOfBibitem
\bibitem[Bonaf\'e \emph{et~al.}(2018)Bonaf\'e, Hernández, Aradi, Frauenheim,
  and S\'anchez]{Bonafe2018JPCL}
F.~P. Bonaf\'e, F.~J. Hernández, B.~Aradi, T.~Frauenheim and C.~G. S\'anchez,
  \emph{J. Phys. Chem. Lett.}, 2018, \textbf{9}, 4355--4359\relax
\mciteBstWouldAddEndPuncttrue
\mciteSetBstMidEndSepPunct{\mcitedefaultmidpunct}
{\mcitedefaultendpunct}{\mcitedefaultseppunct}\relax
\EndOfBibitem
\bibitem[Gastegger and Marquetand(2015)]{Gastegger2015JCTC}
M.~Gastegger and P.~Marquetand, \emph{J. Chem. Theory Comput.}, 2015,
  \textbf{11}, 2187--2198\relax
\mciteBstWouldAddEndPuncttrue
\mciteSetBstMidEndSepPunct{\mcitedefaultmidpunct}
{\mcitedefaultendpunct}{\mcitedefaultseppunct}\relax
\EndOfBibitem
\bibitem[Akimov(2018)]{Akimov2018JPCL}
A.~V. Akimov, \emph{J. Phys. Chem. Lett.}, 2018, \textbf{9}, 6096--6102\relax
\mciteBstWouldAddEndPuncttrue
\mciteSetBstMidEndSepPunct{\mcitedefaultmidpunct}
{\mcitedefaultendpunct}{\mcitedefaultseppunct}\relax
\EndOfBibitem
\bibitem[Ryabinkin \emph{et~al.}(2017)Ryabinkin, Joubert-Doriol, and
  Izmaylov]{Ryabinkin2017ACR}
I.~G. Ryabinkin, L.~Joubert-Doriol and A.~F. Izmaylov, \emph{Acc. Chem. Res.},
  2017, \textbf{50}, 1785--1793\relax
\mciteBstWouldAddEndPuncttrue
\mciteSetBstMidEndSepPunct{\mcitedefaultmidpunct}
{\mcitedefaultendpunct}{\mcitedefaultseppunct}\relax
\EndOfBibitem
\bibitem[Matsika and Krause(2011)]{Matsika2011ARPC}
S.~Matsika and P.~Krause, \emph{Annu. Rev. Phys. Chem.}, 2011, \textbf{62},
  621--643\relax
\mciteBstWouldAddEndPuncttrue
\mciteSetBstMidEndSepPunct{\mcitedefaultmidpunct}
{\mcitedefaultendpunct}{\mcitedefaultseppunct}\relax
\EndOfBibitem
\bibitem[Oloyede \emph{et~al.}(2006)Oloyede, Mil\'{}nikov, and
  Nakamura]{Oloyede2006JCP}
P.~Oloyede, G.~Mil\'{}nikov and H.~Nakamura, \emph{J. Chem. Phys.}, 2006,
  \textbf{124}, 144110\relax
\mciteBstWouldAddEndPuncttrue
\mciteSetBstMidEndSepPunct{\mcitedefaultmidpunct}
{\mcitedefaultendpunct}{\mcitedefaultseppunct}\relax
\EndOfBibitem
\bibitem[Ishida \emph{et~al.}(2017)Ishida, Nanbu, and Nakamura]{Ishida2017IRPC}
T.~Ishida, S.~Nanbu and H.~Nakamura, \emph{Int. Rev. Phys. Chem.}, 2017,
  \textbf{36}, 229--286\relax
\mciteBstWouldAddEndPuncttrue
\mciteSetBstMidEndSepPunct{\mcitedefaultmidpunct}
{\mcitedefaultendpunct}{\mcitedefaultseppunct}\relax
\EndOfBibitem
\bibitem[Zhu \emph{et~al.}(2002)Zhu, Kamisaka, and Nakamura]{Zhu2002JCP}
C.~Zhu, H.~Kamisaka and H.~Nakamura, \emph{J. Chem. Phys.}, 2002, \textbf{116},
  3234--3247\relax
\mciteBstWouldAddEndPuncttrue
\mciteSetBstMidEndSepPunct{\mcitedefaultmidpunct}
{\mcitedefaultendpunct}{\mcitedefaultseppunct}\relax
\EndOfBibitem
\bibitem[Ishida \emph{et~al.}(2009)Ishida, Nanbu, and Nakamura]{Ishida2009JPCA}
T.~Ishida, S.~Nanbu and H.~Nakamura, \emph{J. Phys. Chem. A}, 2009,
  \textbf{113}, 4356--4366\relax
\mciteBstWouldAddEndPuncttrue
\mciteSetBstMidEndSepPunct{\mcitedefaultmidpunct}
{\mcitedefaultendpunct}{\mcitedefaultseppunct}\relax
\EndOfBibitem
\bibitem[Gao \emph{et~al.}(2012)Gao, Li, Zhang, and Han]{Gao2012JCP}
A.-H. Gao, B.~Li, P.-Y. Zhang and K.-L. Han, \emph{J. Chem. Phys.}, 2012,
  \textbf{137}, 204305\relax
\mciteBstWouldAddEndPuncttrue
\mciteSetBstMidEndSepPunct{\mcitedefaultmidpunct}
{\mcitedefaultendpunct}{\mcitedefaultseppunct}\relax
\EndOfBibitem
\bibitem[Yu \emph{et~al.}(2014)Yu, Xu, Lei, Zhu, and Wen]{Yu2014PCCP}
L.~Yu, C.~Xu, Y.~Lei, C.~Zhu and Z.~Wen, \emph{Phys. Chem. Chem. Phys.}, 2014,
  \textbf{16}, 25883--25895\relax
\mciteBstWouldAddEndPuncttrue
\mciteSetBstMidEndSepPunct{\mcitedefaultmidpunct}
{\mcitedefaultendpunct}{\mcitedefaultseppunct}\relax
\EndOfBibitem
\bibitem[Mai \emph{et~al.}(2015)Mai, Marquetand, and Gonz\'alez]{Mai2015IJQC}
S.~Mai, P.~Marquetand and L.~Gonz\'alez, \emph{Int. J. Quantum Chem.}, 2015,
  \textbf{115}, 1215--1231\relax
\mciteBstWouldAddEndPuncttrue
\mciteSetBstMidEndSepPunct{\mcitedefaultmidpunct}
{\mcitedefaultendpunct}{\mcitedefaultseppunct}\relax
\EndOfBibitem
\bibitem[Plasser \emph{et~al.}(2016)Plasser, Ruckenbauer, Mai, Oppel,
  Marquetand, and Gonz\'alez]{Plasser2016JCTC}
F.~Plasser, M.~Ruckenbauer, S.~Mai, M.~Oppel, P.~Marquetand and L.~Gonz\'alez,
  \emph{J. Chem. Theory Comput.}, 2016, \textbf{12}, 1207\relax
\mciteBstWouldAddEndPuncttrue
\mciteSetBstMidEndSepPunct{\mcitedefaultmidpunct}
{\mcitedefaultendpunct}{\mcitedefaultseppunct}\relax
\EndOfBibitem
\bibitem[Lischka \emph{et~al.}(2001)Lischka, Shepard, Pitzer, Shavitt, Dallos,
  M\"uller, Szalay, Seth, Kedziora, Yabushita, and Zhang]{Lischka2001PCCP}
H.~Lischka, R.~Shepard, R.~M. Pitzer, I.~Shavitt, M.~Dallos, T.~M\"uller, P.~G.
  Szalay, M.~Seth, G.~S. Kedziora, S.~Yabushita and Z.~Zhang, \emph{Phys. Chem.
  Chem. Phys.}, 2001, \textbf{3}, 664--673\relax
\mciteBstWouldAddEndPuncttrue
\mciteSetBstMidEndSepPunct{\mcitedefaultmidpunct}
{\mcitedefaultendpunct}{\mcitedefaultseppunct}\relax
\EndOfBibitem
\bibitem[van~der Walt \emph{et~al.}(2011)van~der Walt, Colbert, and
  Varoquaux]{Walt2011CSE}
S.~van~der Walt, S.~C. Colbert and G.~Varoquaux, \emph{Comput. Sci. Eng.},
  2011, \textbf{13}, 22--30\relax
\mciteBstWouldAddEndPuncttrue
\mciteSetBstMidEndSepPunct{\mcitedefaultmidpunct}
{\mcitedefaultendpunct}{\mcitedefaultseppunct}\relax
\EndOfBibitem
\bibitem[{Theano Development Team}(2016)]{TDT2016a}
{Theano Development Team}, \emph{arXiv}, 2016,  1605.02688 [cs.SC]\relax
\mciteBstWouldAddEndPuncttrue
\mciteSetBstMidEndSepPunct{\mcitedefaultmidpunct}
{\mcitedefaultendpunct}{\mcitedefaultseppunct}\relax
\EndOfBibitem
\bibitem[Neese(2012)]{Neese2012WCMS}
F.~Neese, \emph{WIREs Comput. Mol. Sci.}, 2012, \textbf{2}, 73--78\relax
\mciteBstWouldAddEndPuncttrue
\mciteSetBstMidEndSepPunct{\mcitedefaultmidpunct}
{\mcitedefaultendpunct}{\mcitedefaultseppunct}\relax
\EndOfBibitem
\bibitem[Levine \emph{et~al.}(2007)Levine, Coe, and Martinez]{Levine2007JPCB}
B.~G. Levine, J.~D. Coe and T.~J. Martinez, \emph{J. Phys. Chem. B}, 2007,
  \textbf{112}, 405--413\relax
\mciteBstWouldAddEndPuncttrue
\mciteSetBstMidEndSepPunct{\mcitedefaultmidpunct}
{\mcitedefaultendpunct}{\mcitedefaultseppunct}\relax
\EndOfBibitem
\bibitem[Bearpark \emph{et~al.}(1994)Bearpark, Robb, and
  Schlegel]{Bearpark1994CPL}
M.~J. Bearpark, M.~A. Robb and H.~B. Schlegel, \emph{Chem. Phys. Lett.}, 1994,
  \textbf{223}, 269\relax
\mciteBstWouldAddEndPuncttrue
\mciteSetBstMidEndSepPunct{\mcitedefaultmidpunct}
{\mcitedefaultendpunct}{\mcitedefaultseppunct}\relax
\EndOfBibitem
\bibitem[Barbatti \emph{et~al.}(2006)Barbatti, Aquino, and
  Lischka]{Barbatti2006MP}
M.~Barbatti, A.~J.~A. Aquino and H.~Lischka, \emph{Mol. Phys.}, 2006,
  \textbf{104}, 1053--1060\relax
\mciteBstWouldAddEndPuncttrue
\mciteSetBstMidEndSepPunct{\mcitedefaultmidpunct}
{\mcitedefaultendpunct}{\mcitedefaultseppunct}\relax
\EndOfBibitem
\bibitem[Herbst \emph{et~al.}(2002)Herbst, Heyne, and Diller]{Herbst2002S}
J.~Herbst, K.~Heyne and R.~Diller, \emph{Science}, 2002, \textbf{297},
  822--825\relax
\mciteBstWouldAddEndPuncttrue
\mciteSetBstMidEndSepPunct{\mcitedefaultmidpunct}
{\mcitedefaultendpunct}{\mcitedefaultseppunct}\relax
\EndOfBibitem
\bibitem[Yarkony(2005)]{Yarkony2005JCP}
D.~R. Yarkony, \emph{J. Chem. Phys.}, 2005, \textbf{123}, 204101\relax
\mciteBstWouldAddEndPuncttrue
\mciteSetBstMidEndSepPunct{\mcitedefaultmidpunct}
{\mcitedefaultendpunct}{\mcitedefaultseppunct}\relax
\EndOfBibitem
\bibitem[Hudock \emph{et~al.}(2007)Hudock, Levine, Thompson, Satzger, Townsend,
  Gador, Ullrich, Stolow, and Mart{\'i}nez]{Hudock2007JPCA}
H.~R. Hudock, B.~G. Levine, A.~L. Thompson, H.~Satzger, D.~Townsend, N.~Gador,
  S.~Ullrich, A.~Stolow and T.~J. Mart{\'i}nez, \emph{J. Phys. Chem. A}, 2007,
  \textbf{111}, 8500--8508\relax
\mciteBstWouldAddEndPuncttrue
\mciteSetBstMidEndSepPunct{\mcitedefaultmidpunct}
{\mcitedefaultendpunct}{\mcitedefaultseppunct}\relax
\EndOfBibitem
\bibitem[Glorot and Bengio(2010)]{Glorot2010}
X.~Glorot and Y.~Bengio, Proceedings of the Thirteenth International Conference
  on Artificial Intelligence and Statistics, Chia Laguna Resort, Sardinia,
  Italy, 2010, pp. 249--256\relax
\mciteBstWouldAddEndPuncttrue
\mciteSetBstMidEndSepPunct{\mcitedefaultmidpunct}
{\mcitedefaultendpunct}{\mcitedefaultseppunct}\relax
\EndOfBibitem
\bibitem[Kingma and Ba(2014)]{Adam2014}
D.~P. Kingma and J.~Ba, ICLR 2015, 2014\relax
\mciteBstWouldAddEndPuncttrue
\mciteSetBstMidEndSepPunct{\mcitedefaultmidpunct}
{\mcitedefaultendpunct}{\mcitedefaultseppunct}\relax
\EndOfBibitem
\bibitem[Wigner(1932)]{Wigner1932PR}
E.~Wigner, \emph{Phys. Rev.}, 1932, \textbf{40}, 749--750\relax
\mciteBstWouldAddEndPuncttrue
\mciteSetBstMidEndSepPunct{\mcitedefaultmidpunct}
{\mcitedefaultendpunct}{\mcitedefaultseppunct}\relax
\EndOfBibitem
\bibitem[Rupp \emph{et~al.}(2012)Rupp, Tkatchenko, M{\"u}ller, and von
  Lilienfeld]{Rupp2012PRL}
M.~Rupp, A.~Tkatchenko, K.-R. M{\"u}ller and O.~A. von Lilienfeld, \emph{Phys.
  Rev. Lett.}, 2012, \textbf{108}, 058301\relax
\mciteBstWouldAddEndPuncttrue
\mciteSetBstMidEndSepPunct{\mcitedefaultmidpunct}
{\mcitedefaultendpunct}{\mcitedefaultseppunct}\relax
\EndOfBibitem
\bibitem[Pedregosa \emph{et~al.}(2011)Pedregosa, Varoquaux, Gramfort, Michel,
  Thirion, Grisel, Blondel, Prettenhofer, Weiss, Dubourg, Vanderplas, Passos,
  Cournapeau, Brucher, Perrot, and Duchesnay]{scikit-learn}
F.~Pedregosa, G.~Varoquaux, A.~Gramfort, V.~Michel, B.~Thirion, O.~Grisel,
  M.~Blondel, P.~Prettenhofer, R.~Weiss, V.~Dubourg, J.~Vanderplas, A.~Passos,
  D.~Cournapeau, M.~Brucher, M.~Perrot and E.~Duchesnay, \emph{Journal of
  Machine Learning Research}, 2011, \textbf{12}, 2825--2830\relax
\mciteBstWouldAddEndPuncttrue
\mciteSetBstMidEndSepPunct{\mcitedefaultmidpunct}
{\mcitedefaultendpunct}{\mcitedefaultseppunct}\relax
\EndOfBibitem
\bibitem[Pukrittayakamee \emph{et~al.}(2009)Pukrittayakamee, Malshe, Hagan,
  Raff, Narulkar, Bukkapatnum, and Komanduri]{Pukrittayakamee2009JCP}
A.~Pukrittayakamee, M.~Malshe, M.~Hagan, L.~M. Raff, R.~Narulkar,
  S.~Bukkapatnum and R.~Komanduri, \emph{J. Chem. Phys.}, 2009, \textbf{130},
  134101\relax
\mciteBstWouldAddEndPuncttrue
\mciteSetBstMidEndSepPunct{\mcitedefaultmidpunct}
{\mcitedefaultendpunct}{\mcitedefaultseppunct}\relax
\EndOfBibitem
\bibitem[C.~Thompson and A.~Collins(1997)]{Thompson1997JCSFT}
K.~C.~Thompson and M.~A.~Collins, \emph{J. Chem. Soc.{,} Faraday Trans.}, 1997,
  \textbf{93}, 871--878\relax
\mciteBstWouldAddEndPuncttrue
\mciteSetBstMidEndSepPunct{\mcitedefaultmidpunct}
{\mcitedefaultendpunct}{\mcitedefaultseppunct}\relax
\EndOfBibitem
\bibitem[O'Boyle \emph{et~al.}(2011)O'Boyle, Banck, James, Morley,
  Vandermeersch, and Hutchison]{OpenBabel2011}
N.~M. O'Boyle, M.~Banck, C.~A. James, C.~Morley, T.~Vandermeersch and G.~R.
  Hutchison, \emph{J. Cheminf.}, 2011, \textbf{3}, 33\relax
\mciteBstWouldAddEndPuncttrue
\mciteSetBstMidEndSepPunct{\mcitedefaultmidpunct}
{\mcitedefaultendpunct}{\mcitedefaultseppunct}\relax
\EndOfBibitem
\bibitem[Helgaker \emph{et~al.}(1997)Helgaker, Jensen, P.~J{\o}rgensen, Ruud,
  {\AA}gren, Andersen, Bak, Bakken, Christiansen, Dahle, Dalskov, Enevoldsen,
  Heiberg, Hettema, Jonsson, Kirpekar, Kobayashi, Koch, Mikkelsen, Norman,
  Packer, Saue, Taylor, and Vahtras]{Helgaker1997}
T.~Helgaker, H.~J.~A. Jensen, J.~O. P.~J{\o}rgensen, K.~Ruud, H.~{\AA}gren,
  T.~Andersen, K.~L. Bak, V.~Bakken, O.~Christiansen, P.~Dahle, E.~K. Dalskov,
  T.~Enevoldsen, H.~Heiberg, H.~Hettema, D.~Jonsson, S.~Kirpekar, R.~Kobayashi,
  H.~Koch, K.~V. Mikkelsen, P.~Norman, M.~J. Packer, T.~Saue, P.~R. Taylor and
  O.~Vahtras, \emph{DALTON, An Ab Initio Electronic Structure Program, Release
  1.0}, 1997\relax
\mciteBstWouldAddEndPuncttrue
\mciteSetBstMidEndSepPunct{\mcitedefaultmidpunct}
{\mcitedefaultendpunct}{\mcitedefaultseppunct}\relax
\EndOfBibitem
\bibitem[Mai \emph{et~al.}(2014)Mai, Richter, Ruckenbauer, Oppel, Marquetand,
  and Gonz\'alez]{sharc-md}
S.~Mai, M.~Richter, M.~Ruckenbauer, M.~Oppel, P.~Marquetand and L.~Gonz\'alez,
  \emph{SHARC: Surface Hopping Including Arbitrary Couplings -- Program Package
  for Non-Adiabatic Dynamics}, sharc-md.org, 2014\relax
\mciteBstWouldAddEndPuncttrue
\mciteSetBstMidEndSepPunct{\mcitedefaultmidpunct}
{\mcitedefaultendpunct}{\mcitedefaultseppunct}\relax
\EndOfBibitem
\bibitem[Sellner \emph{et~al.}(2013)Sellner, Barbatti, M\"{u}ller, Domcke, and
  Lischka]{Sellner2013MP}
B.~Sellner, M.~Barbatti, T.~M\"{u}ller, W.~Domcke and H.~Lischka,
  \emph{Molecular Physics}, 2013, \textbf{111}, 2439--2450\relax
\mciteBstWouldAddEndPuncttrue
\mciteSetBstMidEndSepPunct{\mcitedefaultmidpunct}
{\mcitedefaultendpunct}{\mcitedefaultseppunct}\relax
\EndOfBibitem
\end{mcitethebibliography}
\end{document}